\pdfoutput=1
\documentclass[aps,prd,a4paper,11pt,notitlepage,superscriptaddress,longbibliography,nofootinbib]{revtex4-1}
\usepackage{microtype}
\usepackage{graphicx}
\graphicspath{{./figures/}}
\usepackage{subfigure}

\usepackage{mathtools}
\usepackage{amssymb}
\DeclareMathOperator{\sgn}{sgn}

\usepackage[dvipsnames]{xcolor}

\usepackage[utf8]{inputenc}
\usepackage[T1]{fontenc}
\usepackage[pdfusetitle,bookmarksopen,colorlinks,allcolors=blue]{hyperref}

\begin{document}
\title{Signum-Gordon spectral mass from nonlinear Fourier mode mixing}

% LTeX: enabled=false
\author{J. S.~Streibel}
\email{jstreibel@gmail.com }
\affiliation{Departamento de Física, Universidade Federal de Santa Catarina,\\
	Campus Trindade, 88040-900, Florianópolis, Brazil}
% LTeX: enabled=true
\author{P.~Klimas}
\email{pawel.klimas@ufsc.br}
\affiliation{Departamento de Física, Universidade Federal de Santa Catarina,\\
	Campus Trindade, 88040-900, Florianópolis, Brazil}

\begin{abstract}
We investigate the emergence of a spectral mass in the signum-Gordon model, a nonlinear field theory characterized by a non-analytic, V-shaped potential where standard perturbative mass definitions are inapplicable. By analyzing the evolution of monochromatic wave trains, we identify two distinct dynamical regimes governed by the relationship between the wave's amplitude and its wavenumber. In the nonlinear regime, the model exhibits nonlinear Fourier mode mixing, where the potential's lack of analyticity acts as a source that populates higher-order harmonics. Using two complementary numerical methods -- tracking frequency distributions from initial wavenumbers and measuring spatial responses to boundary signals -- we construct comprehensive dispersion maps in energy-momentum space. Our results demonstrate that the signum-Gordon field effectively mimics a massive theory. Specifically, we show that a particular initial wave amplitude induces a spectral mass of unity, perfectly matching the behavior of the massive Klein-Gordon equation and providing a robust framework for quantifying mass in non-analytic scalar models.
 \end{abstract}

\maketitle

\section{Introduction}
 Scalar field models with nonlinear field equations find broad applications in various fields of physics, ranging from particle physics to cosmology \cite{Rubakov:2002fi,Arod2003PatternsOS}. Nonlinearity profoundly impacts the properties of physical models, including topological features \cite{Manton:2004tk, Shnir:2018yzp}, integrability \cite{Alvarez:1997ma, Alvarez2009INTEGRABLETA, Dunajski:2010zz}, and the compactness of field solutions \cite{PhysRevLett.70.564, PhysRevE.57.2320, Arodz:2002yt}. The fundamental dynamics of field configurations, such as internal modes and energy transfer between these modes, continues to be a subject of intense investigation in field theory \cite{Adam:2023kel, Adam:2023qgx, Adam:2023qgx, Oles:2023ujf, Blaschke:2024uec, AlonsoIzquierdo:2024nbn, Marjaneh:2024yec, Simas:2024qdq, Alonso-Izquierdo:2024uei}. In particular, the concept of mass plays a crucial role in these studies. Even within classical field theory, the meaning of mass can vary depending on the specific context. For instance, the mass attributed to field configurations like kinks, Q-balls, and oscillons typically represents their {\it energy}. The mass parameter can also be introduced through the field's self-interaction potential, $V(\phi)$. Specifically, the {\it perturbative mass} is defined as the second derivative of the potential evaluated at its minimum, $\phi_\text{min}$:
 \begin{equation}\label{pert_mass}
m^2_0:=\frac{d^2V}{d\phi^2}\Big|_{\phi_\text{min}}.
\end{equation}
For instance, this mass plays a crucial role in determining the dispersion relation between wavenumber $k$ and frequency $\omega$ in the Klein-Gordon (KG) model. Consequently, a nonzero perturbative mass implies a discrepancy between the phase velocity and the group velocity, both of which are functions of the wavenumber, $k$. In certain contexts within field theory, the perturbative mass alone is insufficient to fully understand the system's behavior. For instance, to explain the existence of oscillons in theories with $m^2=0$, the concept of an {\it effective mass} (or smeared mass) was introduced in \cite{Dorey:2023sjh}. This mass is defined through the integral
\begin{equation}
m^2_{\text{eff}}:=\int_{-\infty}^{\infty}\;d\phi\, w_{\sigma}(\phi)\frac{d^2V}{d\phi^2}
\end{equation}
where the weight function $w_{\sigma}(\phi)$ isolates the behavior of $\frac{d^2V}{d\phi^2}$ near the potential minimum. This mass incorporates the effects of field oscillations around the vacuum state. The weight function can be chosen with a degree of flexibility without significantly affecting the qualitative behavior of the perturbative mass. One possible choice for the weight function is the Heaviside step function:
\[
w_{\sigma}(\phi)=\frac{1}{2\sigma}\theta\Big(1-\frac{|\phi-\phi_\text{min}|}{\sigma}\Big).
\]

The problem of mass must be treated with caution in models with non-analytic potentials, where the classical definition of the perturbative mass is inapplicable due to the inconsistent behavior of side derivatives at the potential minimum \cite{Arodz:2005gz}. In such cases, the perturbative mass can only be represented by a generalized function, such as the Dirac delta distribution.

%%%%%%%%%%%%%%%%%%%%%%%%%%%%%%%%%%%%%%%%%%%%%%%%%%%%%%%%%

A perturbative mass, similar to its role in the Klein-Gordon (KG) model, appears in the dispersion relation that determines the allowed wavenumbers $k$ for a wave signal of a given frequency $\omega$. For the KG equation, this relation is derived directly from the Fourier transform of the linear field equation, resulting in the algebraic condition:
\[
\omega^2=k^2+m_0^2,
\]
where $m_0$ represents the bare mass of the scalar field as defined in Eq.~\eqref{pert_mass}. This relation implies a lower frequency bound -- or mass gap -- at $\omega = m_0$, below which plane-wave solutions become evanescent.

In the context of the signum-Gordon (SG) model \cite{Arodz:2007ek}, however, the situation is more nuanced due to the non-smooth nature of the V-shaped potential $V(\phi) = |\phi|$. A natural question arises: what is the form of the dispersion relation for such a model, and what "mass" would appear in this relation, given that the standard perturbative mass is not well-defined at the vacuum?

This paper is devoted to the determination of a characteristic mass derived from the dispersion relation, which we refer to as the {\it spectral mass} to reflect the physical context of its derivation. To address this, we excite the signum-Gordon vacuum in two distinct ways: through specific initial field configurations and via signals generated during the evolution. We then perform a numerical Fourier analysis of the resulting field evolution, which provides access to the dispersion curves (or dispersion maps) used to estimate the spectral mass parameter.

The emergence of such a mass is strictly related to the coupling between different modes. The nonlinearity of the SG model is responsible for nonlinear Fourier mode mixing, which allows for the excitation of higher modes during the evolution, even if only a single mode was initially present. In the first part of this work, we estimate the leading correction to the spectral mass by analyzing this nonlinear mixing. Finally, the characteristic mass is validated through a comparative analysis with the nonlinear Klein-Gordon model.

%%%%%%%%%%%%%%%%%%%%%%%%%%%%%%%%%%%%%%%%%%%%%%%%%%%%%%%%%

The paper is organized as follows. In Section \ref{sec:properties}, we briefly review the most characteristic properties of the signum-Gordon (SG) model. Subsequently, in Section \ref{sec:smasslessandmassive}, we discuss the massless and massive sectors of the model as a function of the incident wave amplitude. In particular, this section details the nonlinear Fourier mode mixing and calculates the leading contribution to the spectral mass. The final part of the section is devoted to the presentation of numerical results and the analysis of dispersion maps.

\section{Characteristic features of the signum-Gordon model}\label{sec:properties}

The signum-Gordon (SG) model is a piecewise linear field theory with the potential $V(\phi)=\lambda_0 |\phi|$, where the scalar field $\phi$ can be either real-valued or complex. The nonlinear character of the model arises from the non-analytic behavior of the potential at its minimum, $\phi_0=0$. The derivative term is considered a standard quadratic function of the field's four-gradient, resulting in the following Lagrangian density:
\begin{equation}\label{lagrangian}
\mathcal{L}=\frac{1}{2}(\partial_\mu\phi)^2 - |\phi|
\end{equation}
where the coupling constant $\lambda_0$ has been absorbed into a redefinition of the dimensionless coordinates $x^{\mu}$, $\mu=0,1,2,3$. Here $\phi\in\mathbb{R}$. The Euler-Lagrange equation, given by
\begin{equation}\label{sgeq}
\partial_{\mu}\partial^{\mu}\phi+\sgn\phi=0
\end{equation}
contains the signum function, which gives rise to the model's name. 

The first important fact is that $\phi_0=0$ is a vacuum solution of the model and a minimizer of the energy-momentum tensor. This solution does not automatically satisfy the Euler-Lagrange equations derived for the sectors $\frac{dV}{d\phi}=\pm 1$. To simplify the analysis and include this physical state within the class of solutions, we define the signum function at $\phi=0$ to be zero.

The second characteristic property is the absence of a linear sector for small-amplitude field excitations. Unlike many field models where the potential can be approximated by a quadratic function near the minimum, the SG potential retains its form regardless of the field's amplitude. The SG model breaks the harmonic oscillator paradigm. Its nonlinearity cannot be scaled away.

Furthermore, the SG equation is invariant under the scaling transformation $x^{\mu}\to \lambda x^{\mu}$, $\phi\to\lambda^{-2}\phi$. In other words, if $\phi(t,\vec x)$ is a solution to \eqref{sgeq},  then 
\begin{equation}\label{eq:scaling}
\phi_{\lambda}(t,\vec x)=\lambda^{-2}\phi(\lambda t,\lambda \vec x)
\end{equation}
is also a solution. This scaling transformation is a universal property of the model. While it is exact for the SG model, it is approximate for models with non-analytic potentials that exhibit behavior similar to $|\phi|$ near their minima. For instance, small-amplitude perturbations around the vacua $\phi_\text{vac}=\pm1$ in a model with the double-well potential $V(\phi)=\frac{1}{2}|\phi^2-1|$ are governed by the SG equation.

The SG model (and related models featuring non-analytic potentials) is relevant in diverse physical contexts. Its genesis lies in the discretized version, which historically first appeared in the description of a chain of coupled pendulums \cite{PhysRevLett.70.564}. Subsequently, a chain of impacting pendulums was proposed as a physical system capable of supporting compact topological kinks \cite{Arodz:2002yt, Arodz:2003ab, Arodz:2005cm}. Another intriguing possibility is the emergence of the model from other physical models through symmetry reduction. This is the case of submodels of the Skyrme model, where a 1+1 dimensional effective SG model was obtained as a result of such a reduction \cite{Adam:2017srx}. 

The SG model is a crucial framework for investigating the hyper-massive behavior of field excitations. This behavior has been observed in recent numerical simulations of compact kink and oscillon scattering \cite{Hahne:2023dic, Hahne:2019ela, Hahne:2022wyl}. This hyper-massive behavior is linked to the threshold force effect, a characteristic of V-shaped potentials. 
A clear example of this effect can be seen in a system of inverted coupled pendulums constrained by rigid barriers in a gravitational field \cite{Arodz:2002yt}. If these pendulums are initially at rest at an angle $\phi_0$ to the vertical, lifting them infinitesimally requires a finite force. As a result, the propagation of impulses is suppressed compared to standard models like the KG model, where infinitesimal displacements can be achieved with infinitesimal forces. Consequently, the SG model emits less radiation than the KG model.  Moreover, the emitted radiation is discrete, taking the form of compact oscillons, whose exact shape is known for the SG model. Following \cite{Dorey:2023sjh} and \cite{Hahne:2024qby}, we find that the square of the perturbative mass for the signum-Gordon model is $m_0=2\delta(\phi)$, which results in an effective mass of
\begin{equation}\label{sgpeffmass}
m^2_\text{eff}=\int_{-\infty}^{\infty}2\delta (\phi)w_{\sigma}(\phi)=\frac{1}{\sigma}.
\end{equation}
To physically interpret this mass, we first need to identify solutions that will provide meaning to the parameter $\sigma$. This is achieved by considering the propagation of a wave train and analyzing its dispersion relation.

%%%%%%%%%%%%%%%%%%%%%%%%%%%%%%%%%%%%%%%%%%%%%%%%%%%%%%%
%%%%%%%%%%%%%%%%%%%%%%%%%%%%%%%%%%%%%%%%%%%%%%%%%%%%%%%
\section{Evolution of wave trains}\label{sec:smasslessandmassive}
\subsection{Massive and massless regime}
%%%%%%%%%%%%%%%%%%%%%%%%%%%%%%%%%%%%%%%%%%%%%%%%%%%%%%%

A wave train, represented in (1+1) dimensions by the monochromatic plane wave expression:
\begin{equation}
	\label{eq:ic_eigen}
	\varphi(t,x) = A_0 \cos(k_0x-\omega_0 t)
\end{equation}
would exhibit differing propagation characteristics within a massless compared to a massive field theory model. Here $\omega_0$ denotes the angular frequency, $k_0$  is the wavenumber, and $A_0$ represents the wave amplitude. The relationship connecting $k_0$ and $\omega_0$ is defined as the dispersion relation. In the elementary case of the free-wave model and the Klein-Gordon model, the expression \eqref{eq:ic_eigen} constitutes a solution, with the only constraint on the wave train being imposed on $k_0$ and $\omega_0$ through the dispersion relation. The wave amplitude $A_0$ remains an arbitrary parameter.

This situation, however, changes significantly for nonlinear field models. Notably, in the signum-Gordon (SG) model, the initial amplitude $A_0$ plays a crucial role in determining the behavior of the resulting solution. It is evident that expression \eqref{eq:ic_eigen} does not constitute a solution to the governing equation:
\begin{equation}\label{eq:phi}
-\partial^2_t\phi=-\partial^2_x\phi+V'(\phi),
\end{equation}
where $V'(\phi)={\rm sgn}(\phi)$ is a piecewise constant term. Therefore, the expression \eqref{eq:ic_eigen} can only serve as an initial field configuration.

A fundamental distinction arises because for expression \eqref{eq:ic_eigen} the signum term $\sgn(\varphi)$ is bounded to $\pm 1$, while the second-derivative terms ($\partial^2_t\varphi$ and $\partial^2_x\varphi$) are proportional to the unconstrained amplitude $A_0$. Consequently, the significance of the nonlinear term $V'(\phi)$ and its effect on the solution depend substantially on the magnitude of $A_0$.

To illustrate this dependence, we substitute expression \eqref{eq:ic_eigen} into the dynamic balance equation:
\[
	\omega_0^2\varphi = k_0^2\varphi+V'(\varphi).
\]
 When the solution's amplitude is small relative to the scale of the signum term, the SG nonlinearity exerts a considerable effect on the solution dynamics. Conversely, for exceptionally large amplitudes, the signum term's influence is reduced to that of a minor perturbation.

This derived relationship would be equivalent to a standard dispersion relation if $\varphi$ could be factored out. This factorization is permissible in the Klein-Gordon (KG) equation, where $V'(\phi)$ is proportional to $\phi$. However, due to the inherently nonlinear nature of $V'(\phi)$ in the SG equation, such a relationship can only be achieved approximately when the signum term is negligible compared to the spatial second-derivative term, $-\partial^2_x\varphi$. This condition is expressed as:
\begin{equation}\label{eq:free_cond}
	|V'(\varphi)|\ll |k_0^2\varphi|.
\end{equation}
%%%%%%%%%%%%%%%%%%%%%%%%%%%%%%%%%%%%%%%%%%%%%%%%%%%%%%%%
%%%%%%%%%%%%%%%%%%%%%%%% FIGURE_1 %%%%%%%%%%%%%%%%%%%%%%%%%%
\begin{figure}[h!]
\centering
\subfigure[]{\includegraphics[width=0.4\textwidth, height=0.25\textwidth]{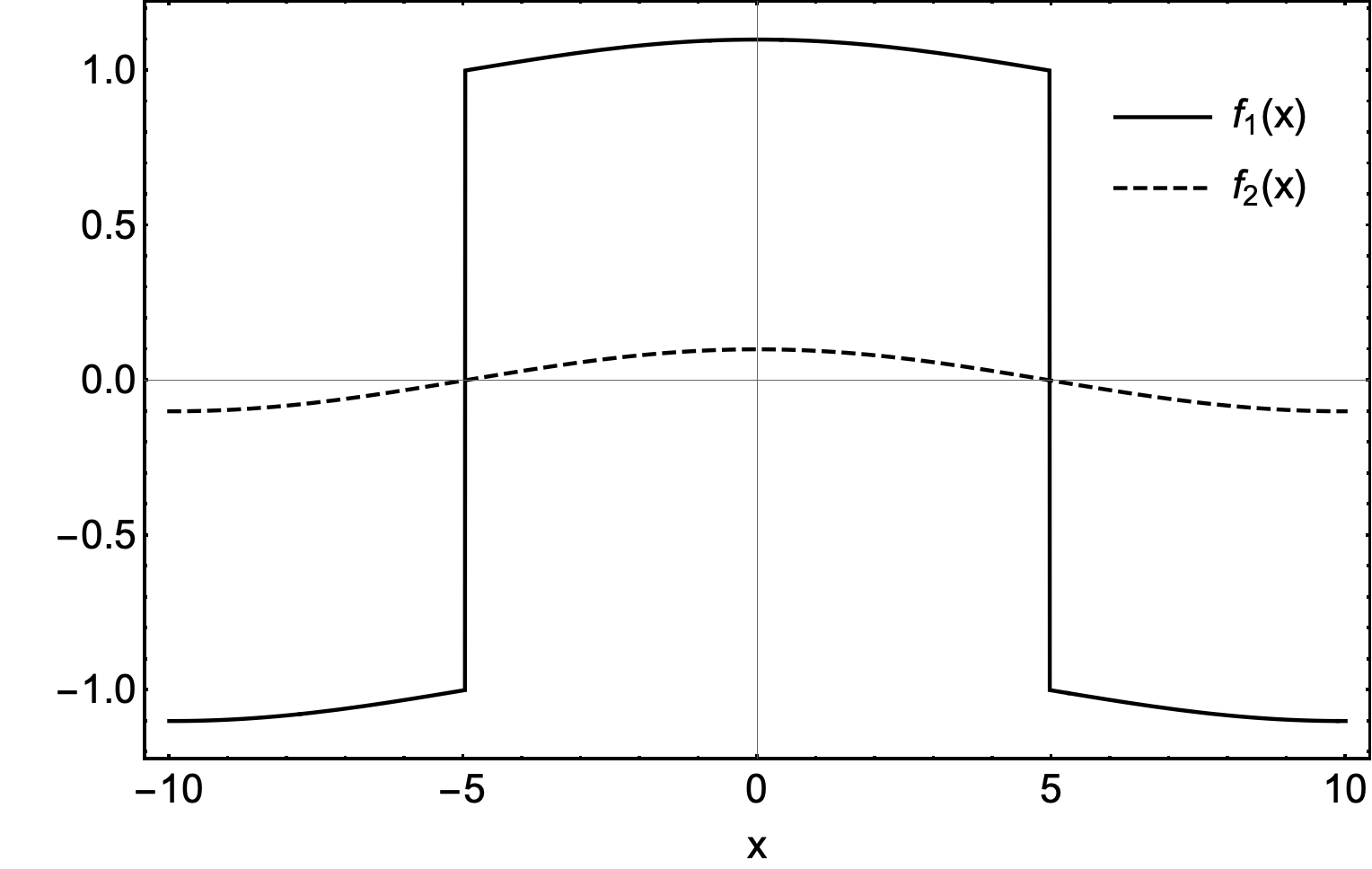}}\hskip0.5cm
\subfigure[]{\includegraphics[width=0.4\textwidth,height=0.25\textwidth]{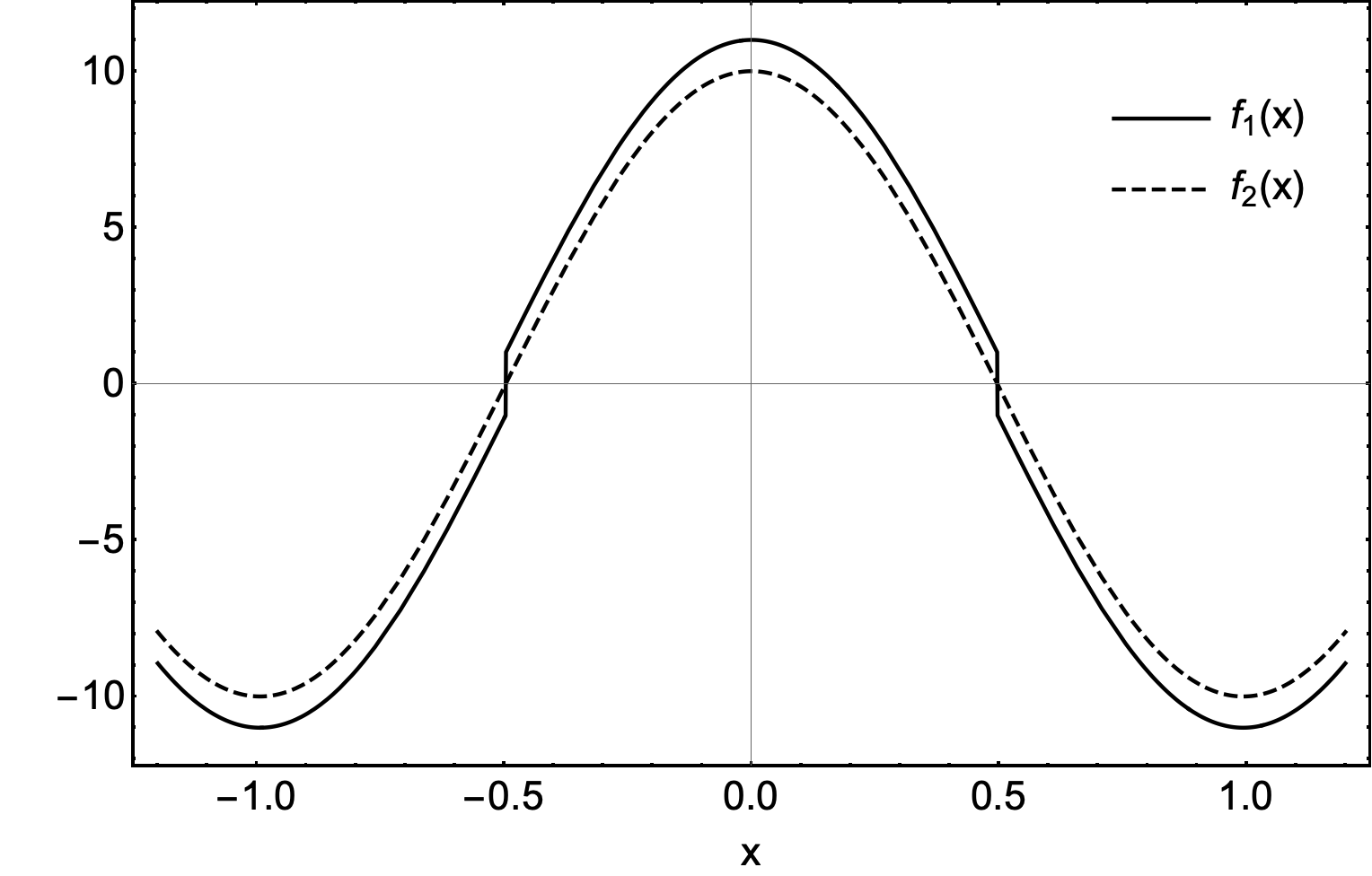}} 
\caption{Comparative Analysis of the field equation's right-hand side. The function $f(x):=-\partial^2_x\varphi+V'(\varphi)$ is plotted at $t=0$ for the field configuration \eqref{eq:ic_eigen} ($A_0=1$). Solid lines show the SG expression, $f_1(x)=A_0k_0^2\cos(k_0x)+\sgn(\cos(k_0x))$, and dashed lines show the free wave case, $f_2(x)=A_0k_0^2\cos(k_0x)$. (a) $A_0k_0^2=10^{-1}$ (Low-Amplitude/Wavenumber). (b) $A_0k_0^2=10$ (High-Amplitude/Wavenumber). The enhanced alignment between the two curves for greater values of $A_0k_0^2$ confirms the convergence to the massless limit (free propagation with $\omega_0\simeq\pm k_0$). Note that the linear behavior, $f_2(x)$, already dominates when $A_0k_0^2=10$.}
\label{fig:sg_rhs}
\end{figure}
%%%%%%%%%%%%%%%%%%%%%%%%%%%%%%%%%%%%%%%%%%%%%%%%%%%%%%%%
%%%%%%%%%%%%%%%%%%%%%%%%%%%%%%%%%%%%%%%%%%%%%%%%%%%%%%%%

Under this condition, the approximate dispersion relation closely resembles that of the free wave equation, specifically $\omega_0^2=k_0^2$. The condition \eqref{eq:free_cond} is relaxed in regions where the field, $\varphi$, is negligible. In these areas, the potential term,  $V'(\varphi)$, may dominate over $k_0^2\varphi$. However, in most other regions, Equation \eqref{eq:phi} effectively reduces to the massless wave equation. In this high-amplitude limit, the condition
 \begin{equation}\label{eq:condamplitude}
 A_0k_0^2\gg 1
 \end{equation}
 provides a more suitable representation of the relevant physical behavior than Equation \eqref{eq:phi} itself, as it directly captures the dynamics.
 It is important to note that the criterion \eqref{eq:condamplitude} remains invariant under the global scaling transformation \eqref{eq:scaling}, which is also a symmetry of the SG equation. Under this transformation, $A_0$ scales as $\lambda^{-2}A_0$ and $k_0$ scales as $\lambda k_0$, thereby preserving the quantity $A_0k_0^2$.
 
Figure \ref{fig:sg_rhs} presents a comparative analysis of the right-hand sides of Equation \eqref{eq:phi} for two distinct cases: the SG equation (where condition \eqref{eq:condamplitude} is not satisfied) and the free wave equation ($V=0$, where the condition holds). Both terms are evaluated at $t=0$ using the initial field configuration provided by Equation \eqref{eq:ic_eigen}.

%%%%%%%%%%%%%%%%%%%%%%%%%%%%%%%%%%%%%%%%%%%%%%%%%%%%%%%
%%%%%%%%%%%%%%%%%%%%%%%%%%%%%%%%%%%%%%%%%%%%%%%%%%%%%%%
\subsection{Fourier modes evolution}
The distinction between the massless and (ultra-)massive regimes of the SG model is particularly evident in the evolution of its Fourier modes. To illustrate this phenomenon, the SG equation was numerically integrated using the initial field configuration provided by Equation \eqref{eq:ic_eigen} at $t=0$. The parameters were established as $\omega_0=k_0=10\pi$, and the system was evolved up to a final time of $t=30$. Periodic boundary conditions were imposed on a spatial domain of length $L=1$, which was discretized using $N=1000$ grid points. The time evolution utilized a 4th-order Runge-Kutta method with a fixed timestep of $\delta t=10^{-4}$.

\begin{figure}[h!]
\centering
\subfigure[]{\includegraphics[width=0.4\textwidth, height=0.2\textwidth]{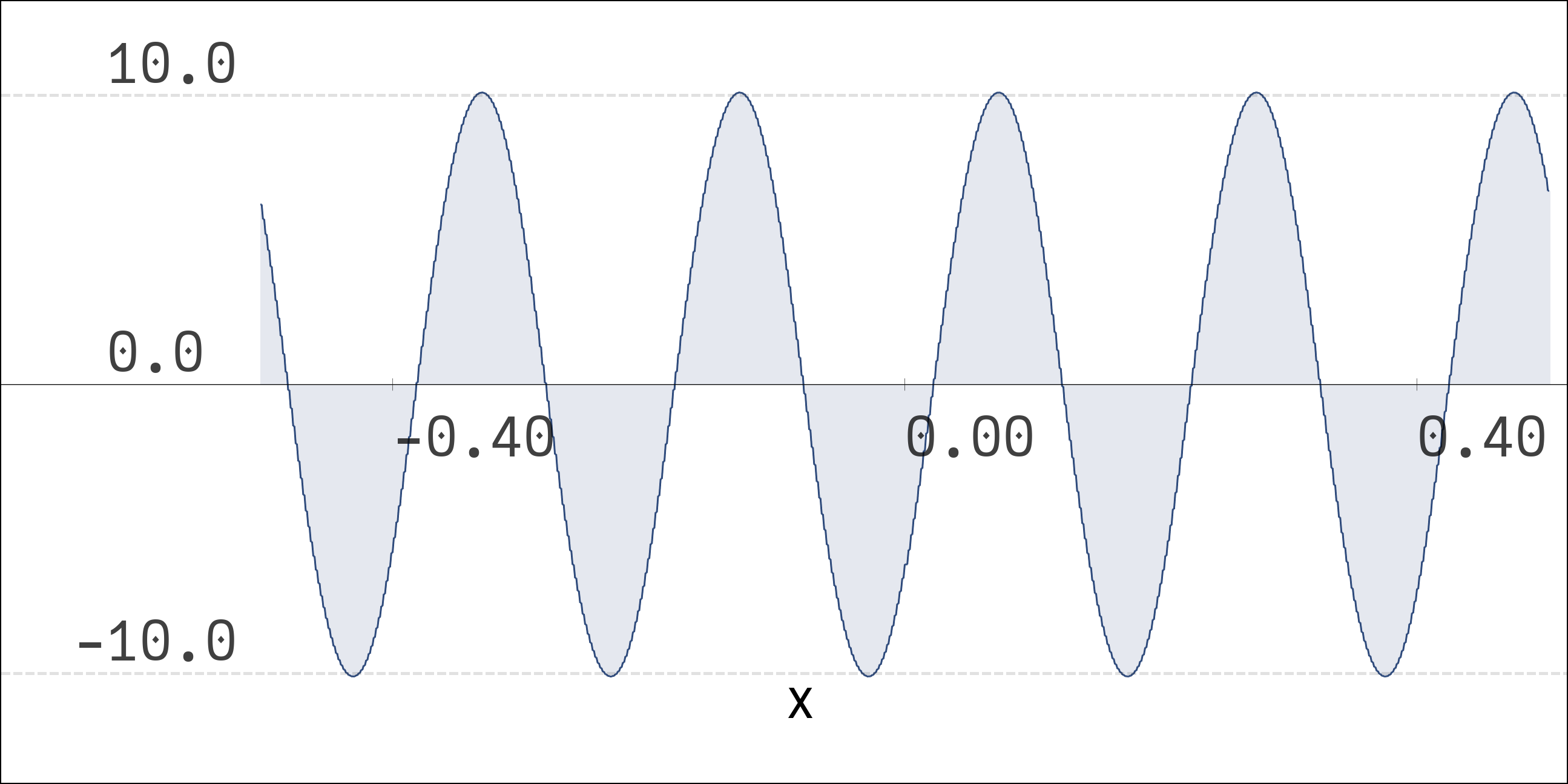}}\hskip0.5cm
\subfigure[]{\includegraphics[width=0.4\textwidth,height=0.2\textwidth]{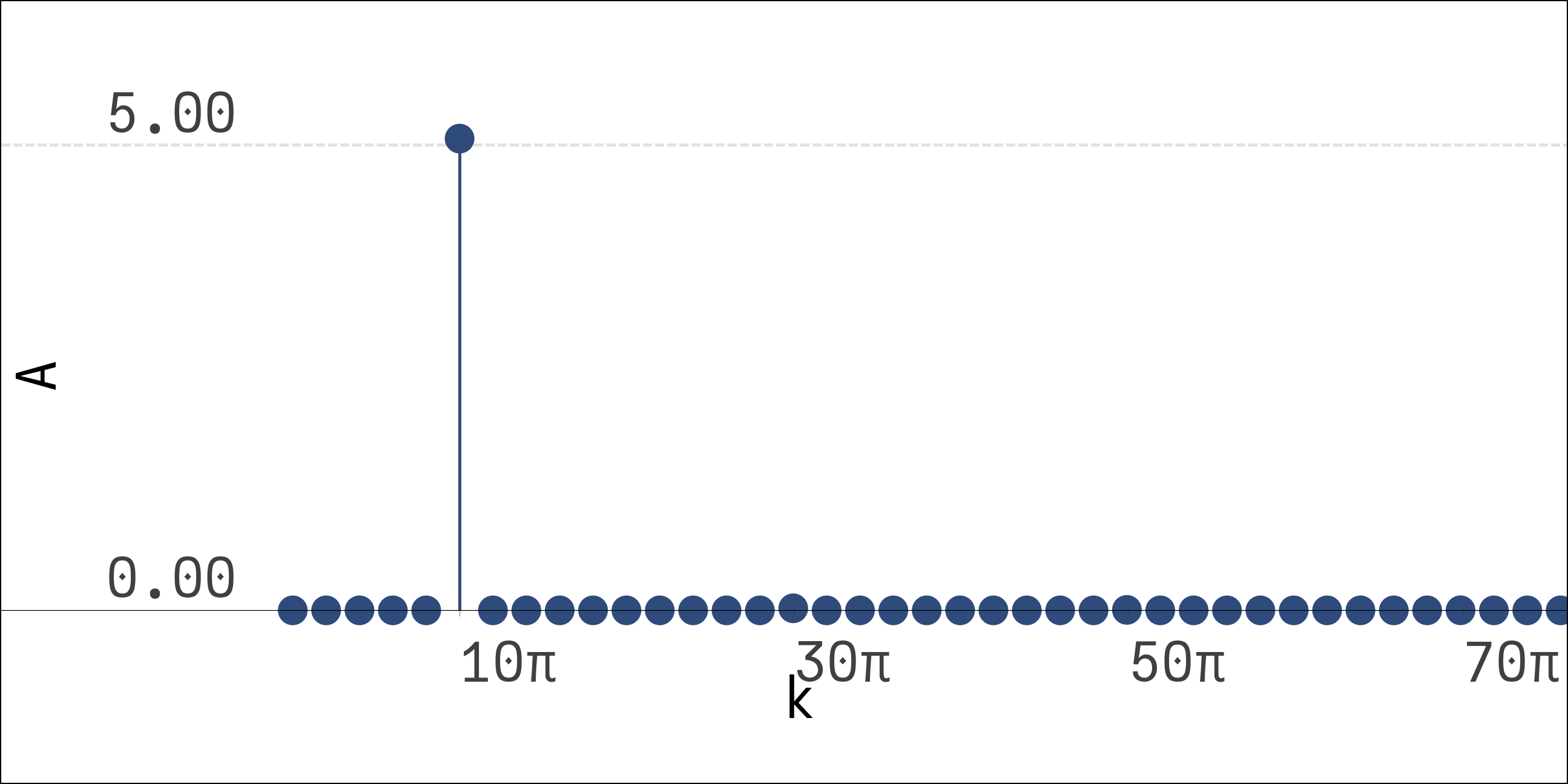}} 
\subfigure[]{\includegraphics[width=0.4\textwidth, height=0.2\textwidth]{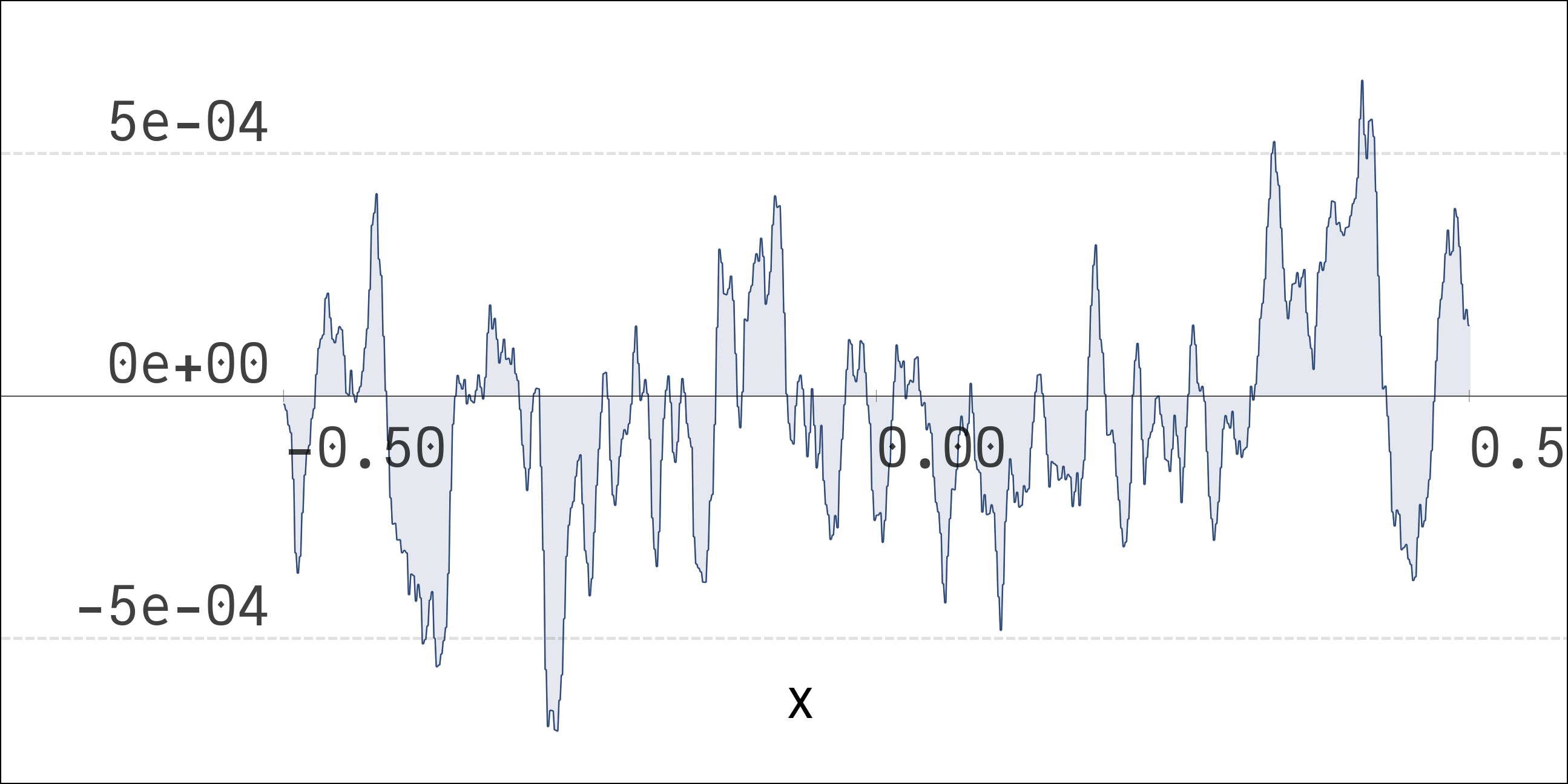}}\hskip0.5cm
\subfigure[]{\includegraphics[width=0.4\textwidth,height=0.2\textwidth]{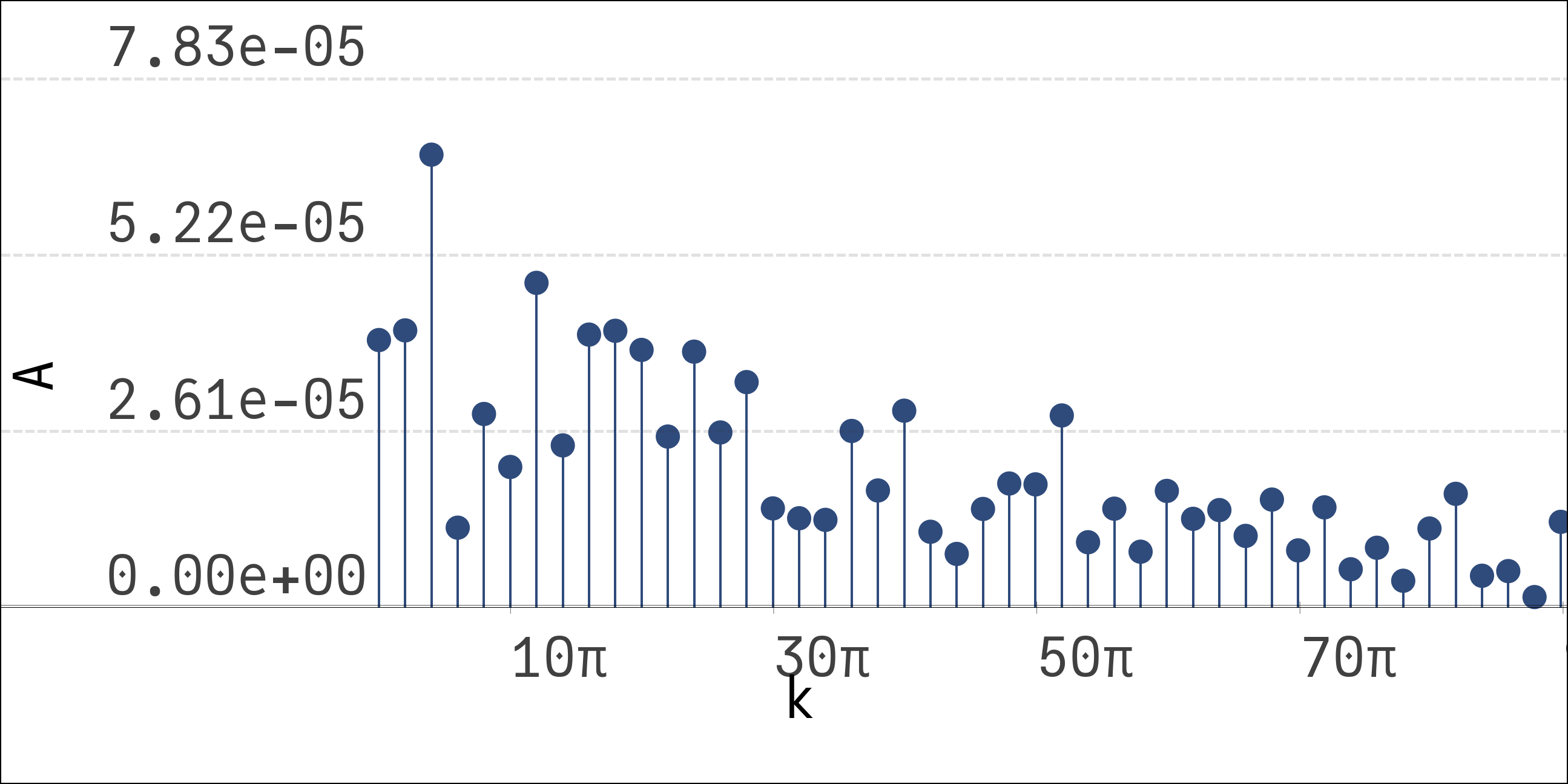}} 
\caption{Field and Fourier amplitudes for two regimes. Final field profiles $\phi(x, t=30)$ and Fourier mode amplitudes $A(k)$ resulting from the numerical solution of the SG equation, initialized by Equation \eqref{eq:ic_eigen}. (a) and (b) Field value and spectrum, respectively, for the free-wave (massless) regime ($A_0k_0^2=10^4$). (c) and (d) Corresponding results for the ultra-massive regime ($A_0k_0^2=1$). The transition from single-mode dominance at high $A_0k_0^2$ to multi-mode excitation at low $A_0k_0^2$ is clearly demonstrated.}
\label{fig:sg-freeandultramassive}
\end{figure}
\begin{figure}[h!]
\centering
\subfigure[]{\includegraphics[width=0.4\textwidth]{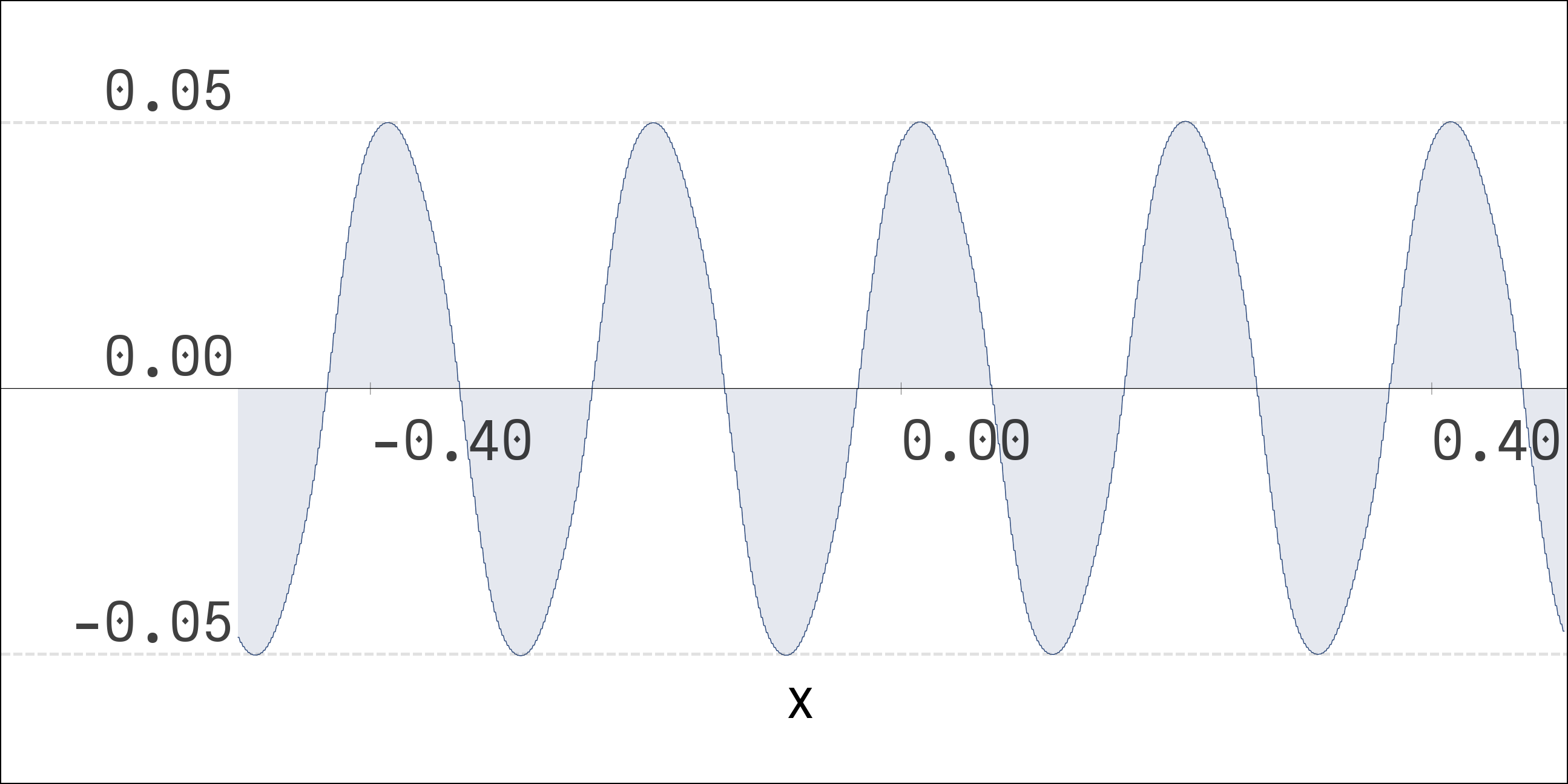}}\hskip0.5cm
\subfigure[]{\includegraphics[width=0.4\textwidth]{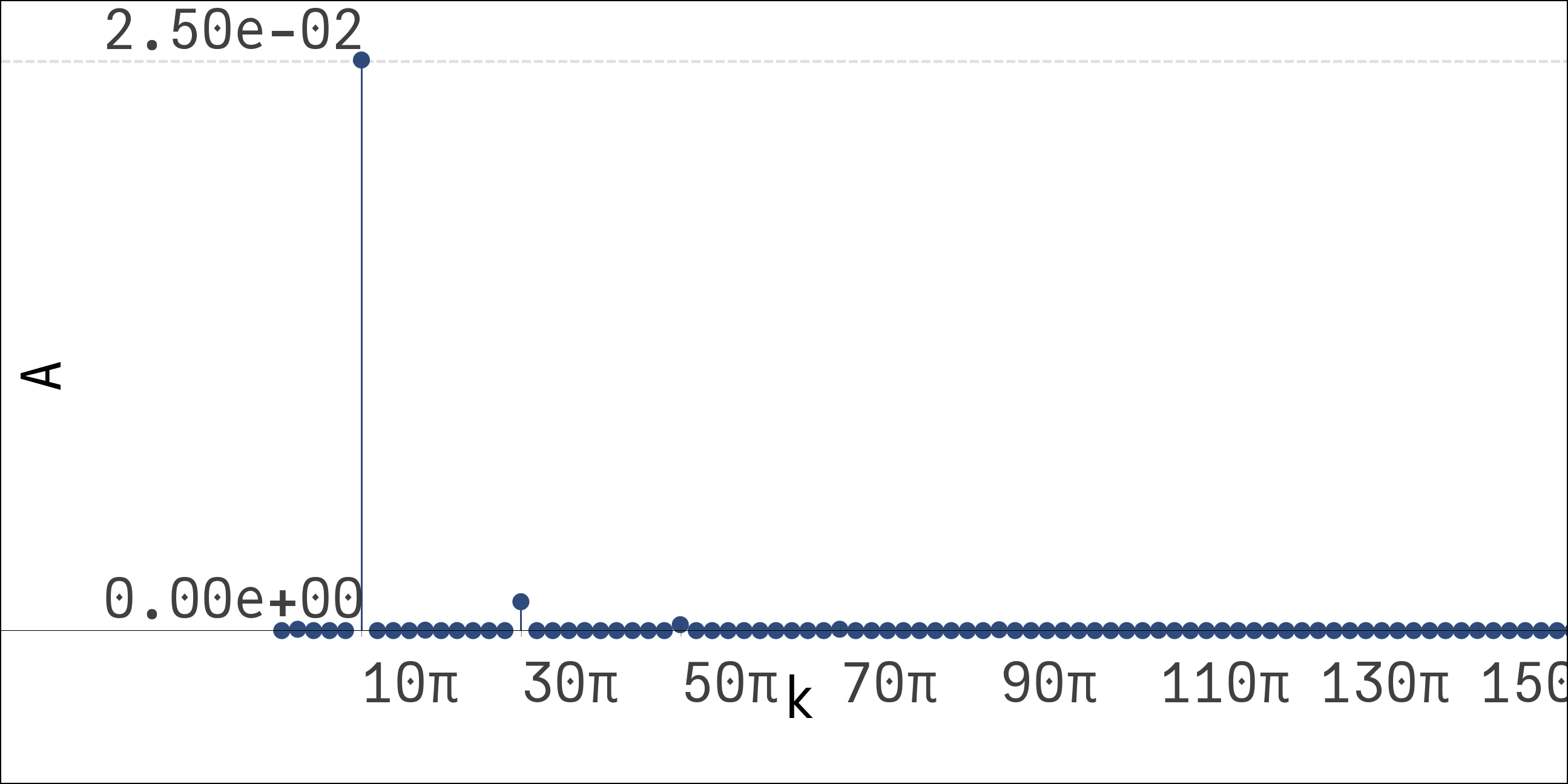}} 
\subfigure[]{\includegraphics[width=0.4\textwidth]{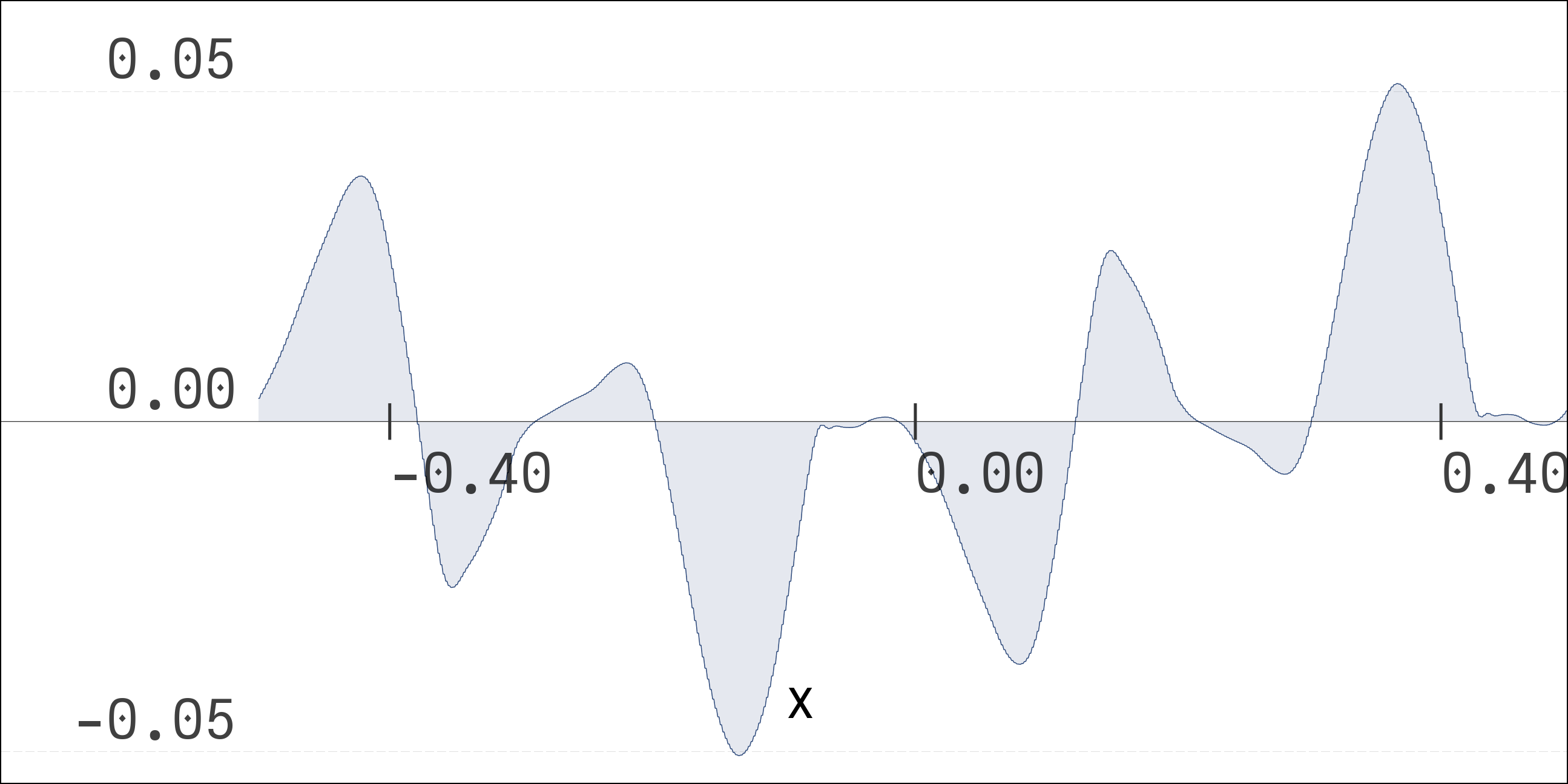}}\hskip0.5cm
\subfigure[]{\includegraphics[width=0.4\textwidth]{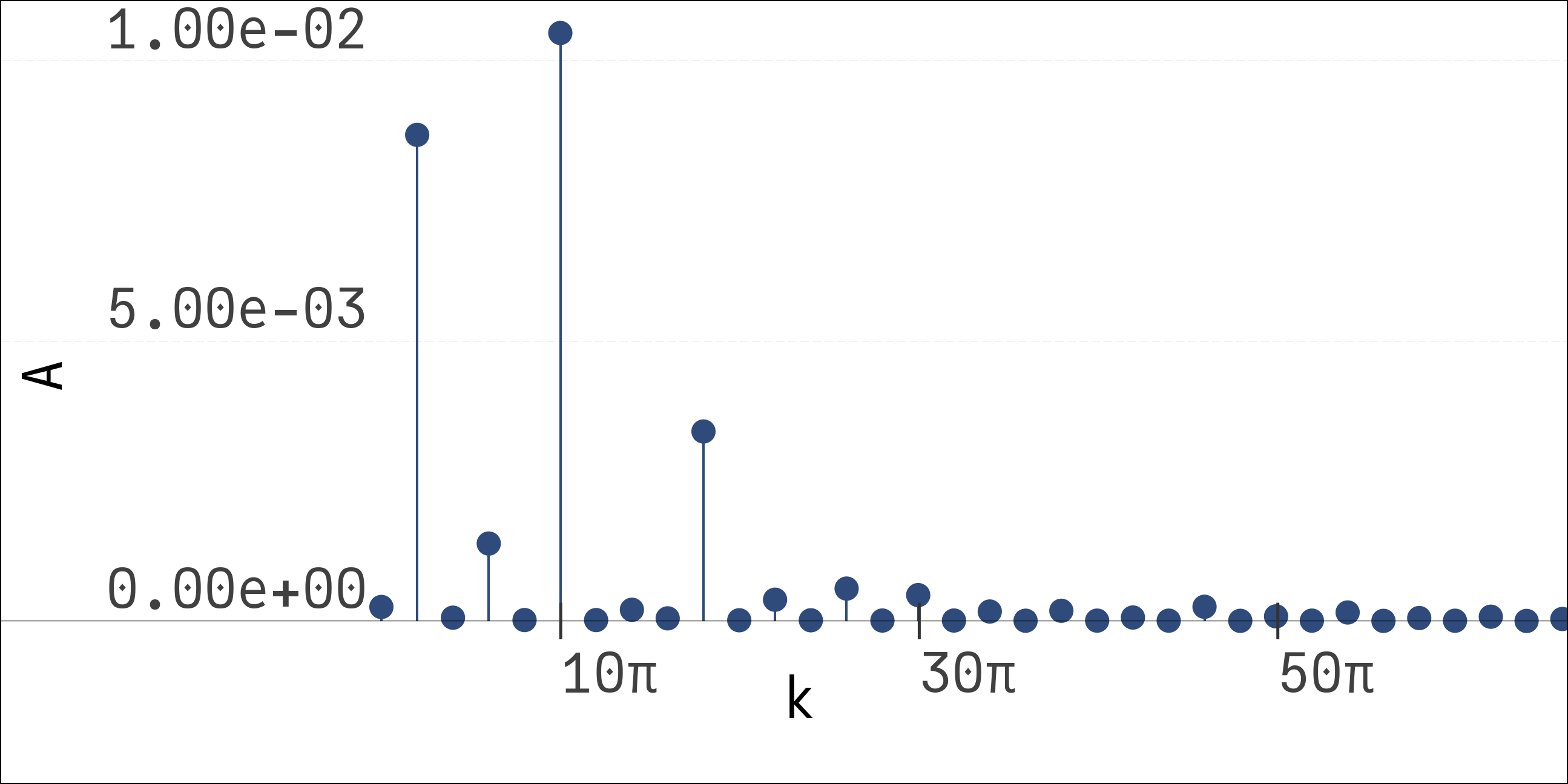}} 
	\caption{Field and Fourier amplitudes at $t=30$ for SG evolution. Field configuration (a, c) and Fourier mode spectrum (b, d) at the final time $t=30$. The simulation was initialized using \eqref{eq:ic_eigen}. Top row (a, b): Results for $A_0k_0^2=50$ (approaching massless limit). Bottom row (c, d): Results for $A_0k_0^2=25$ (transitional regime). The Fourier spectrum in (d) demonstrates nonlinear mode mixing; however, the initial wavenumber $k=10\pi$ persists as the overwhelmingly dominant component.}
\label{fig:sg-massive}
\end{figure}

Figure \ref{fig:sg-freeandultramassive} presents the final field profiles and their corresponding Fourier mode amplitudes at time $t=30$. The amplitude modes, $A(k)$, are plotted as a function of the wavenumber $k$. For cases where the condition $A_0k_0^2\gg 1$ is satisfied (the massless limit), the solution is dominated by a single Fourier mode. Consequently, the field evolution closely approximates the propagation of a monochromatic plane wave governed by the linear wave equation, which is consistent with the approximation developed previously. Conversely, when $A_0k_0^2$ approaches unity (the non-linear, ultra-massive regime), multiple Fourier modes are excited. This excitation leads to a significantly irregular field profile, as visibly demonstrated in Figure \ref{fig:sg-freeandultramassive}(c).

We also examined the case of intermediate values of the parameter $A_0k_0^2$. The results are shown in Figure \ref{fig:sg-massive}. For $A_0k_0^2=50$, the amplitude of the initial mode $k_0$ remains dominant, although the emergence of odd harmonics $k=\{3k_0,5k_0,7k_0,\ldots\}$ becomes noticeable. For a smaller value of the parameter, $A_0k_0^2=25$, a broader spectrum of Fourier modes is activated, encompassing not only multiples of $k_0$ but also modes with wavenumbers  $k<k_0$.

%%%%%%%%%%%%%%%%%%%%%%%%%%%%%%%%%%%%%%%%%%%%%%%%%%%%%%%
%%%%%%%%%%%%%%%%%%%%%%%%%%%%%%%%%%%%%%%%%%%%%%%%%%%%%%%
\subsection{Nonlinear Fourier mode mixing}
\subsubsection{Nonlinear Klein-Gordon equation}
To study the phenomenon of nonlinear mode generation, we consider a Klein-Gordon (KG) equation augmented by an additional quartic term in the field potential. This potential is defined as:
\[
V_3(\phi)=\frac{1}{2}m_0^2\phi^2+\frac{1}{4}\lambda\phi^4,
\]
where the index $3$ highlights the highest power of the nonlinear term present in the resulting field equation:
\[
(\partial^2_t-\partial^2_x)\phi + m_0^2\phi + \lambda\phi^3=0.
\]
This model, commonly known as $\phi^4$ field theory, is notable for possessing kink-type topological solutions. In our specific approach, the quartic term is treated as a nonlinear perturbation of the linear KG model (where $\lambda=0$). Henceforth, we shall refer to this system as the Nonlinear Klein-Gordon (NKG) model.

In the case of the linear Klein-Gordon (KG) equation ($\lambda=0$), the Fourier modes evolve independently. Their dynamics are governed solely by the dispersion relation $\omega_k^2=k^2+m_0^2$, where $\omega_k$ and $k$ are the frequency and wavenumber of each respective mode. The parameter $m_0^2$ corresponds to the perturbative mass defined in Equation \eqref{pert_mass}.

The nonlinear term $\lambda\phi^3$ breaks the independence of the Fourier modes, thereby creating inter-modal coupling. We utilize the Fourier transform, $\hat\phi(t,k)$, as a mathematical tool to explicitly examine the resulting mode coupling. This allows us to represent the field $\phi(t,x)$ as the inverse Fourier transform:
\begin{equation}\label{eq:phi_linear}
\phi(t,x) = \mathcal{F}^{-1}[\hat{\phi}(t,k)]\equiv\frac{1}{2\pi}\int_{-\infty}^{\infty} dk\,\hat{\phi}(t,k) e^{ikx}.
\end{equation}
Expressions involving powers of the function $\phi(t,x)$ can be represented using the inverse Fourier transform of convolutions. Specifically, the product of two functions, $\phi(t,x)$ and $\psi(t,x)$, is expressed as:
\begin{align*}
\phi(t,x) \psi(t,x) &=\mathcal{F}^{-1}[\hat \phi(t,k') ]\mathcal{F}^{-1}[\hat \psi(t,k'') ].
\end{align*}
This product is equivalent to the inverse Fourier transform of the convolution of their respective Fourier transforms, as demonstrated by the convolution theorem:
\begin{align*}
\phi(t,x) \psi(t,x) &= \frac{1}{2\pi}\mathcal{F}^{-1}[(\hat\phi * \hat\psi)(t,k)].
\end{align*}
The convolution of the Fourier transforms of the two functions, $\hat\phi(t,k')$ and $\hat\psi(t,k'')$, denoted by $(\hat\phi * \hat\psi)(t,k)$, is formally defined by the integral:
\[
(\hat\phi * \hat\psi)(t,k):=\int_{-\infty}^{\infty} dk' \hat\phi(t,k')\hat \psi(t,k-k').
\]
The preceding concept can be systematically extended to the product of three or more functions. For the third power (or product of three identical functions), the expression becomes:
\begin{equation}\label{eq:phi_cube}
\phi^3(t,x)=\frac{1}{(2\pi)^2}\mathcal{F}^{-1}[(\hat\phi *(\hat\phi *\hat \phi))(t,k)].
\end{equation}
Here, the term $(\hat\phi *(\hat\phi *\hat \phi))(t,k)$ represents the two-fold convolution of the Fourier transform $\hat\phi(t,k)$, which is explicitly defined by the double integral:
\[
(\hat\phi *(\hat\phi *\hat \phi))(t,k)=\int_{\mathbb{R}^2}dk'dk''\;\hat\phi(t,k')\hat\phi(t,k'')\hat\phi(t,k-k'-k'').
\]
This integral generalizes the convolution operation to three functions in the frequency domain, maintaining the relationship with the product in the spatial domain via the inverse Fourier transform.

The nonlinear terms within the NKG equation are responsible for the generation of higher-frequency harmonics. To illustrate this mechanism, consider an initial field configuration restricted to a single fundamental wavenumber, $k_0$, in the Fourier domain:
\begin{equation}\label{eq_initfreq}
\hat\phi(0,k)=\pi\Big(\delta(k-k_0)+\delta(k+k_0)\Big).
\end{equation}
Applying the generalized convolution formula for the third power (as previously established in $\eqref{eq:phi_cube}$) to this initial configuration, the cubic term $\phi^3(0,x)$ yields the expression:
\[
\phi^3(0,x)=\frac{1}{4}\Big(3\cos(k_0 x)+\cos(3k_0x)\Big)=\cos^3(k_0 x)
\]
This resulting expression contains an additional wavenumber, $3k_0$, which indicates the generation of a higher harmonic. This is in direct contrast to the linear term $\phi(0,x)$, which contains only the fundamental mode:
$$\phi(0,x) = \mathcal{F}^{-1}[\hat\phi(0,k)] = \cos(k_0x).$$
Thus, the presence of the cubic nonlinearity directly translates the interaction of the fundamental mode into the creation of frequency components that are integer multiples of the initial wavenumber.

The mechanism for generating higher harmonics, as previously described, can be generalized to the case involving higher powers of the scalar field in the field equation. This generalization yields a more complex spectrum of frequencies within the resulting field configuration. When generalizing the NKG model to incorporate a potential of the form:
\begin{equation}\label{eq:NKGpotential}
	V_{N}(\phi)=\sum_{n=1,3,5,...}^{N} \frac{\lambda_{n}}{n+1} \phi^{n+1}
\end{equation}
we introduce higher-order terms proportional to $\lambda_n\phi^n$ (where $n=1,3,5,\ldots,N$) into the governing field equations.  Consequently, the prescribed initial condition (referenced in Equation \eqref{eq_initfreq}) would generate higher powers of cosine functions. These higher powers, in turn, introduce higher harmonics as expressed by the following trigonometric identity:
\begin{eqnarray} \label{eq:cos_euler_newton}
	\cos^n (k_0x) =
	\left\{\begin{array}{ll}
		 \dfrac{1}{2^{n-1}}\sum_{j=1,3,...}^{n}\text{C}_{\frac{n+j}{2}}^n\,\cos(jk_0x),&n={1,3,5,...} \\ \\
		\dfrac{1}{2^n}\left(1+2\sum_{j=2,4,...}^{n}\text{C}_{\frac{n+j}{2}}^n\,\cos(jk_0x)\right),&n={2,4,6,...}
	\end{array}\right.,
\end{eqnarray}
where $\text{C}_j^n$ represents the binomial coefficient, $C_k^n\equiv\binom{n}{k}=\frac{n!}{k!(n-k)!}$.  It is important to note that nonlinear terms of the form $\phi^n$ with odd $n$ give rise to $n-2$ additional harmonics. Although the case for even $n$ is included in the identity for completeness, the subsequent analysis will be primarily focused on the implications of odd values of $n$.

When examining the temporal evolution of the system of two first-order equations, which is equivalent to the second-order field equation \eqref{eq:phi}, we find that the initial field configuration inherently contains higher harmonic modes that significantly influence the system's subsequent dynamics. To illustrate this, we express the second-order field equation $\eqref{eq:phi}$ as the following coupled system of first-order equations for the field $\phi(t,x)$ and its conjugate momentum $\Pi(t,x)$:
\begin{equation}
	\left\{\begin{array}{lr}
		\Pi(t,x) = \dfrac{\partial\phi(t,x)}{\partial t} \\
		\ \\
		\dfrac{\partial \Pi}{\partial t} = \dfrac{\partial^2\phi(t,x)}{\partial x^2} - V'\big(\phi(t,x)\big)
	\end{array}\right..
\end{equation}
Integrating this system with respect to time yields the integral forms for the fields:
\begin{align}
\phi(t,x) &= \phi(0,x) + \int_0^t \Pi(t',x) dt'\nonumber\\
\Pi(t,x) &= \Pi(0,x) +  \int_0^t\left(\dfrac{\partial^2\phi(t',x)}{\partial x^2}-V'\big(\phi(t',x)\big)\right)dt' .\nonumber
\end{align}
Assuming that the integrands are well-behaved over an infinitesimal time step $\varepsilon$, we can approximate these integrals by using a simple, first-order approximation based on the initial values (analogous to the Euler method):
\begin{align}
 \label{eq:first_step_phi}
	\phi(\varepsilon,x) &= \phi(0,x) + \varepsilon\,\Pi(0,x) 
\\ \label{eq:first_step_pi}
	\Pi(\varepsilon,x) &= \Pi(0,x) + \varepsilon\left(\dfrac{\partial^2\phi(0,x)}{\partial x^2}-V'\big(\phi(0,x)\big)\right).
\end{align}
Equation $\eqref{eq:first_step_pi}$ explicitly shows that the change in the momentum $\Pi$ depends on the spatial derivative $\big(\frac{\partial^2\phi(0,x)}{\partial x^2}\big)$ and the derivative of the potential $\left(-V'\big(\phi(0,x)\big)\right)$, both evaluated at $t=0$. If the initial field $\phi(0,x)$ is composed primarily of a fundamental wavenumber, $k_0$, the subsequent action of the nonlinear term, $V'(\phi(0,x))$, will inherently introduce higher harmonics (i.e., multiples of $k_0$). Consequently, the nonlinear terms present in the original field equation are directly responsible for the phenomenon where energy is transferred between different wavenumbers, which we term {\it nonlinear Fourier mode mixing}.

%%%%%%%%%%%%%%%%%%%%%%%%%%%%%%%%%%%%%%%%%%%%%%%%%%%%%
\subsubsection{The signum-Gordon model}
The SG model is characterized by a potential that is non-analytic, meaning it cannot be accurately represented by a Taylor series expansion around its minimum. This constraint complicates direct comparison with models derived from analytic potentials, such as the NKG model. 

To establish a connection between the SG and NKG models, we analyze the effect of the potential derivative, $V'(\phi)$, when evaluated using a specific initial wave configuration $\eqref{eq:ic_eigen}$.

We consider the linear, infinitesimal time advancement of the field configurations $\phi(\tau_i,x)$ and $\Pi(\tau_i,x)$ using a small time step $\delta\tau$ (where $\tau_i := i\,\delta\tau$), which is equivalent to a first-order Euler numerical scheme. The iterative evolution is defined by:
\[
\left\{\begin{array}{lr}
\phi(\tau_{i+1},x) = \phi(\tau_i,x) + \delta\tau\,\Pi(\tau_i,x) 
\\ \\ 
	\Pi(\tau_{i+1},x) = \Pi(\tau_i,x) + \delta\tau \left[ \dfrac{\partial^2\phi(\tau_i,x)}{\partial x^2} - V'\big(\phi(\tau_i,x)\big) \right] \\
\end{array}\right. .
\]
When the initial configuration $\eqref{eq:ic_eigen}$ (representing a free wave solution) is substituted into this numerical scheme, the terms associated with the potential derivative must vanish for both the SG and NKG equations in the linear regime. This requirement effectively recovers the free wave condition $\eqref{eq:free_cond}$ for the initial step. Consequently, during the initial stages of the temporal evolution, the potential term $V'(\phi)$ acts as a perturbation to the dynamics described by the linear wave equation, which governs the field's underlying temporal evolution.

The fundamental distinction between the NKG and SG models is rooted entirely in their respective field potentials, $V_{N}(\phi)$ and $V_{SG}(\phi)$. To directly compare the dynamical effects of these differences, we employ the nonlinear potential $V_{N}$ within the established perturbative integration framework. This allows us to assess its influence on an initial sinusoidal wave and directly compare these effects to those produced by the SG potential. Crucially, when evaluated on a sinusoidal initial field configuration, both potential derivatives, $V'_{N}(\phi(0,x))$ and $V'_{SG}(\phi(0,x))$, can be represented as Fourier series. These series are composed exclusively of positive odd harmonics of the fundamental wavenumber $k$. This shared harmonic structure facilitates a direct analysis of nonlinear Fourier mode mixing and allows us to quantify the generation of higher harmonics in the temporal evolution of both the NKG and SG models.

For the NKG model, the potential term introduces a first-order perturbation (proportional to the time step, $\delta t$) to the time evolution. This perturbation is directly related to the derivative of the nonlinear potential, $V'_{N}(\phi(0,x))$, evaluated at the initial field configuration $\phi(0,x)$. The initial perturbation in the momentum evolution is given by:
\begin{align}\label{deltatVnl}
	\delta t\,V'_{N}\left(\varphi(0,x)\right) &= \delta t\,\sum_{n=1,3,5,...}^{N}\lambda_{n}\,A_0^n\,\cos^n(k_0x)\nonumber\\
	&= \delta t\;\sum_{j=1,3,5,...}^{N}A_j\,\cos\big(jk_0x\big).
\end{align} 
The first line represents the perturbation using the polynomial expansion of the potential derivative, where $\lambda_n$ are the expansion coefficients. The second line represents the decomposition of this term into a Fourier cosine series of odd harmonics, $jk_0x$. This transformation reveals the instantaneous generation of higher harmonic modes ($j=3, 5, \dots$) due to the nonlinearity. 

The Fourier coefficients $A_j$ for each harmonic mode are determined by the cumulative effect of the original polynomial coefficients $\lambda_n$. Utilizing the identity $\eqref{eq:cos_euler_newton}$ to decompose powers of cosine, the coefficients $A_j$ are expressed as:
\begin{equation} \label{eq:A_j}
	A_j=2\sum_{n=j,j+2,...}^{N}\lambda_{n}\left(\dfrac{A_0}{2}\right)^n\text{C}_{\frac{j+n}{2}}^n.
\end{equation}
This formula demonstrates how the individual nonlinear terms (indexed by $n$) contribute to the amplitude of a specific harmonic (indexed by $j$), thus quantifying the Fourier mode mixing.

To ensure that the infinite series expansion for $V'_{N}(\phi)$ is mathematically sound and that the resulting harmonic amplitudes $A_j$ converge as the truncation limit $N\rightarrow\infty$, we apply the ratio test. The convergence of $A_j$ is determined by the ratio of successive terms in the summation (which differ in $n$ by 2):
\[
	r=\lim\limits_{n\rightarrow\infty} \left|\dfrac{\lambda_{n}(A_0/2)^n\text{C}_{\frac{j+n}{2}}^n}{\lambda_{n-2}(A_0/2)^{n-2}\text{C}_{\frac{j+n-2}{2}}^{n-2}}\right|=A_0^2\lim\limits_{n\rightarrow\infty}\left|\dfrac{\lambda_{n}}{\lambda_{n-2}}\right|.
\]
For the series to converge, we require $r<1$. This leads directly to the amplitude convergence criterion for the initial field amplitude $A_0$:
\begin{equation} \label{eq:convergence_condition}
	\lim\limits_{n\rightarrow\infty}\left|\dfrac{\lambda_{n-2}}{\lambda_n}\right|>A_0^2.
\end{equation}
This condition places an upper bound on the initial amplitude $A_0$ that can be used while maintaining the validity of the NKG polynomial expansion.

In contrast to the complex, amplitude-dependent expansion of the NKG model, the SG perturbation term takes a remarkably simple, non-analytic form. When the SG potential derivative, $V'_{SG}(\varphi)$, is evaluated using the initial sinusoidal wave $\varphi(0,x) = A_0 \cos(kx)$, the result is not a continuous polynomial but a square wave. Specifically, the derivative of the SG potential (proportional to ${\rm sgn}(\varphi)$ for a certain field range) evaluates to the sign function of the cosine wave $V'_{SG}\left(\varphi(0,x)\right) = \text{sgn}\big(\cos(kx)\big)$. This square wave can be precisely represented as a Fourier series, which is an infinite sum of sinusoidal functions containing only odd harmonics of the fundamental wavenumber $k$. The first-order perturbation (proportional to $\delta t$) is thus given by the following Fourier series:
\begin{align}\label{eq:fouriersigncos}
		\delta t\,V'_{SG}\left(\varphi(0,x)\right) &= \delta t\,\text{sgn}\big(\cos(kx)\big)\nonumber\\     	&= \delta t\,	\dfrac{4}{\pi}\sum_{j=1,3,5,...}^{\infty}\dfrac{(-1)^{\frac{j-1}{2}}}{j}\cos\left(j(kx)\right).
\end{align}
This explicit series provides the harmonic amplitudes directly, showing that the SG nonlinearity instantly generates all higher odd harmonics with an amplitude that decreases algebraically as $1/j$.

Both the NKG perturbation \eqref{deltatVnl} and the SG perturbation \eqref{eq:fouriersigncos} are expressed as Fourier series involving the same set of cosine functions, $\cos\left(jkx\right)$, where $j$ is restricted to odd integers. 

To enable a direct quantitative comparison between the analytic NKG model (with its potential convergence issues) and the non-analytic SG model, we employ a regularization scheme. This involves truncating the infinite SG Fourier series $\eqref{eq:fouriersigncos}$ at a finite harmonic index $N$.
By truncating the index $j$ at $N$, we ensure that the two perturbation terms, $\delta t\,V'_{N}\big(\phi(0,x)\big)$ and $\delta t\,V'_{SG}\big(\phi(0,x)\big)$, are represented by an equal number of corresponding cosine functions. This allows us to equate the coefficients $A_j$ of the NKG expansion (from $\eqref{eq:A_j}$) with the known, fixed coefficients of the SG expansion. Equating the $j$-th harmonic amplitudes yields the following system of linear equations for the NKG potential coefficients, $\lambda_n$:
\begin{equation}
	\sum_{n=j,j+2,...}^{N}\lambda_{n}\left(\dfrac{A_0}{2}\right)^n\,\binom{n}{\frac{j+n}{2}} = \dfrac{2}{j\,\pi}(-1)^{\frac{j-1}{2}}\label{eq:sysstemlambda}
\end{equation}
where the index $j=1, 3, 5, \dots$ up to $N$ defines the equations.

The system of equations $\eqref{eq:sysstemlambda}$ can be recast into a standard linear matrix system by introducing the re-indexed variables $\alpha=\frac{j-1}{2}$ and $\beta=\frac{N-1}{2}$ (where $\alpha \in \{0, 1, \ldots, \beta\}$). The system then takes the form: 
\[
\sum_{\gamma=0,1,2,\ldots}^{\beta-\alpha} {\cal A}_{\alpha \gamma}\,\lambda_{2\alpha+2\gamma+1}={\cal B}_{\alpha}, \qquad \alpha\le \beta.
\]
The system's matrix coefficients are explicitly defined as:
\[
{\cal A}_{\alpha \gamma}\equiv\left(\frac{A_0}{2}\right)^{2\alpha+2\gamma+1}\binom{2\alpha+2\gamma+2}{2\alpha+\gamma+1}, \qquad {\cal B}_{\alpha}\equiv\frac{2(-1)^{\alpha}}{(2\alpha+1)\pi}
\]
 Since the upper summation limit for $\gamma$ is $(\beta-\alpha)$, which depends on $\alpha$, the coefficient matrix $\mathcal{A}_{\alpha \gamma}$ forms a triangular matrix. This structure is highly advantageous, as it allows the system to be solved sequentially for the coefficients $\lambda_{2\alpha+1}$ (i.e., $\lambda_1, \lambda_3, \lambda_5, \dots$) through a straightforward process of back-substitution.
 
 The coefficients $\lambda_n$ of the NKG potential, derived by matching the Fourier expansion of the SG perturbation, inherently depend on the truncation parameter $N$ used in the regularization scheme. To explicitly denote this dependency, we introduce the notation $\lambda_n^{(N)}$.
 
For the maximum index $n=N$ (the last coefficient solved for in the triangular system), the solution is found directly:
\[
\lambda_N\equiv\lambda_{2\beta+1}=\frac{{\cal B}_{\beta}}{{\cal A}_{\beta 0}}=(-1)^{\frac{N-1}{2}}\frac{2}{N\pi}\left(\frac{2}{A_0}\right)^N.
\]
More generally, the explicit solution for any odd index $n \le N$ is given by:
\begin{equation} \label{eq:lambda_n}
\lambda_n^{(N)} = (-1)^\frac{n-1}{2}\dfrac{2}{n\pi}\binom{\frac{N+n}{2}}{\frac{N-n}{2}}\left(\dfrac{2}{A_0}\right)^n.
\end{equation}
This formula effectively defines the analytic potential $V_{N}$ whose perturbative effects on the initial sinusoidal wave exactly mimic those of the non-analytic SG potential up to the order $N$.

The derived expression \eqref{eq:lambda_n} must satisfy the amplitude convergence criterion \eqref{eq:convergence_condition} established earlier:
$$
\lim\limits_{n\rightarrow\infty}\left|\dfrac{\lambda_{n-2}}{\lambda_n}\right|>A_0^2.
$$

Substituting the explicit form of $\lambda_n^{(N)}$ into the ratio test (using the relationship $\lambda_{n-2}/\lambda_n$) yields the condition:
$$
\frac{ n^2(n-1)}{(n-2)(N+n)(N-n+2)} > 1.
$$
This condition must hold for the coefficients to converge as $n$ increases for a fixed $N$. In the limit case where $n=N$, considering the ratio of the last two coefficients, $\lambda^{(N)}_{N-2}/\lambda^{(N)}_{N}$, the convergence condition simplifies and is satisfied for any $N>2$. More generally, the condition for general convergence is satisfied for all indices $n \le N$ when $N$ is sufficiently large. This verification confirms that the constructed NKG potential inherits the convergence properties required for an analytic field theory while remaining perturbatively equivalent to the non-analytic SG model in the initial time step.

The lowest-order coefficient, $\lambda_1^{(N)}$ (where the index 1 corresponds to the linear term in the potential derivative), exhibits a clear linear dependence on the truncation parameter $N$. This term is directly identified with the perturbative mass squared ($m_0^2$) of the equivalent NKG model:
\begin{equation}\label{eq:perturbativemass}
m_0^2 \equiv \lambda_1^{(N)} = \frac{2}{A_0\pi}(N+1).
\end{equation}
This formula reveals a crucial dependency: two identical wave trains with the same initial amplitude $A_0$ will exhibit a difference in their perturbative mass based on the parameter $N$. Since $N$ is associated with the inclusion of $\frac{N+1}{2}$ coupling constants $\lambda_n^{(N)}$, the perturbative mass $m_0^2$ is directly proportional to the number of coupling constants used in the NKG approximation. The absolute lowest perturbative mass corresponds to the $N=1$ truncation, where $m_0^2 = \frac{4}{A_0\pi}$.

We can establish a connection between this perturbative mass $m_0^2$ and the intrinsic effective mass $m^2_\text{eff}$ of the SG model $\eqref{sgpeffmass}$. By identifying the wave amplitude parameter $A_0$ with the size of the SG perturbation $\sigma$, the relationship becomes:
\begin{equation}\label{eq:effectivemass}
m^2_\text{eff} = \frac{1}{A_0} = \frac{\pi}{2(N+1)}m_0^2.
\end{equation}
This equation explicitly links the mass scale of the SG model (determined by $A_0$) to the mass derived from the NKG approximation (determined by $N$).

To analyze the behavior of the remaining coupling constants $\lambda_n^{(N)}$, we consider the asymptotic limit as the truncation parameter $N \rightarrow \infty$. We examine the expression $\lim_{N\rightarrow\infty}\frac{2}{N}\ln |\lambda^{(N)}_n|$. By applying Stirling's approximation for the factorials within the combinatorial term $\binom{\frac{N+n}{2}}{\frac{N-n}{2}}$ (specifically: $\ln k!=\left(k+\frac{1}{2}\right)\ln k-k+\frac{1}{2}\ln(2\pi)$), we obtain the asymptotic behavior in terms of the continuous ratio $\nu \equiv n/N$:
\begin{equation}\lim_{N\rightarrow\infty}\frac{2}{N}\ln|\lambda^{(N)}n|=\ln \lambda(\nu)\end{equation}
where the function $\lambda(\nu)$ is defined as:
\begin{equation}\label{eq:lambda_nu}\lambda(\nu) \equiv \dfrac{1+\nu}{1-\nu}\left(\frac{1-\nu^2}{A_0^2\nu^2}\right)^{\nu}.
\end{equation}
This allows us to approximate the general coupling constant $\lambda_{n}^{(N)}$ in the large $N$ limit as:
\begin{equation} \label{eq:lambda_n_lambda_nu}\lambda_{n}^{(N)} \simeq (-1)^\frac{n-1}{2}\left[\lambda\left(\dfrac{n}{N}\right)\right]^{N/2}.
\end{equation}This asymptotic expression is crucial as it reveals the functional dependence of the higher-order coefficients on the ratio of the harmonic index $n$ to the truncation index $N$, effectively describing the profile of the NKG potential in the continuum limit.

%%%%%%%%%%%%%%%%%%

\begin{figure}[h!]
\centering
\subfigure[]{\includegraphics[width=0.45\textwidth]{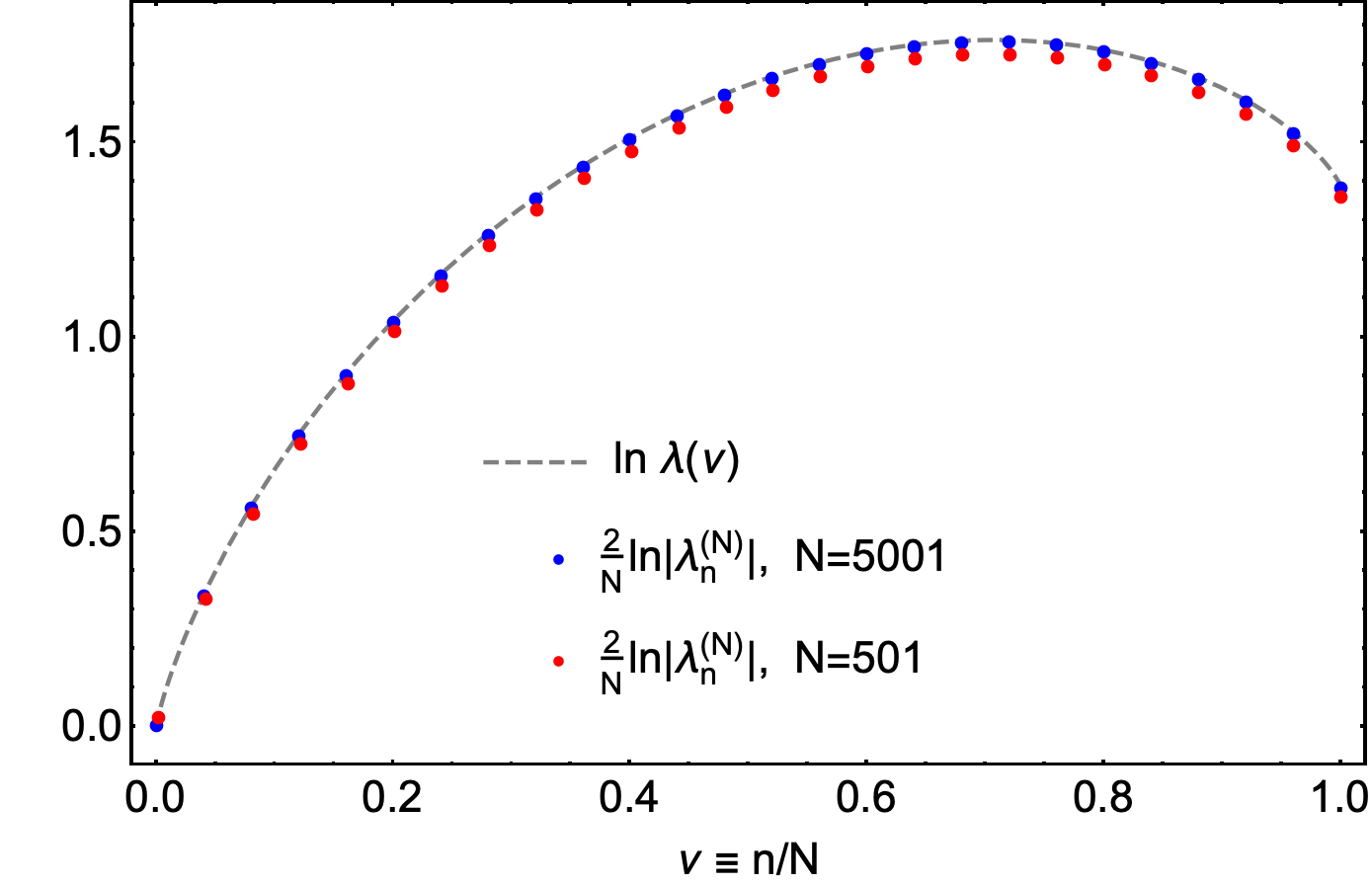}\label{fig:lambda_nu2_comp}}
\subfigure[]{\includegraphics[width=0.45\textwidth]{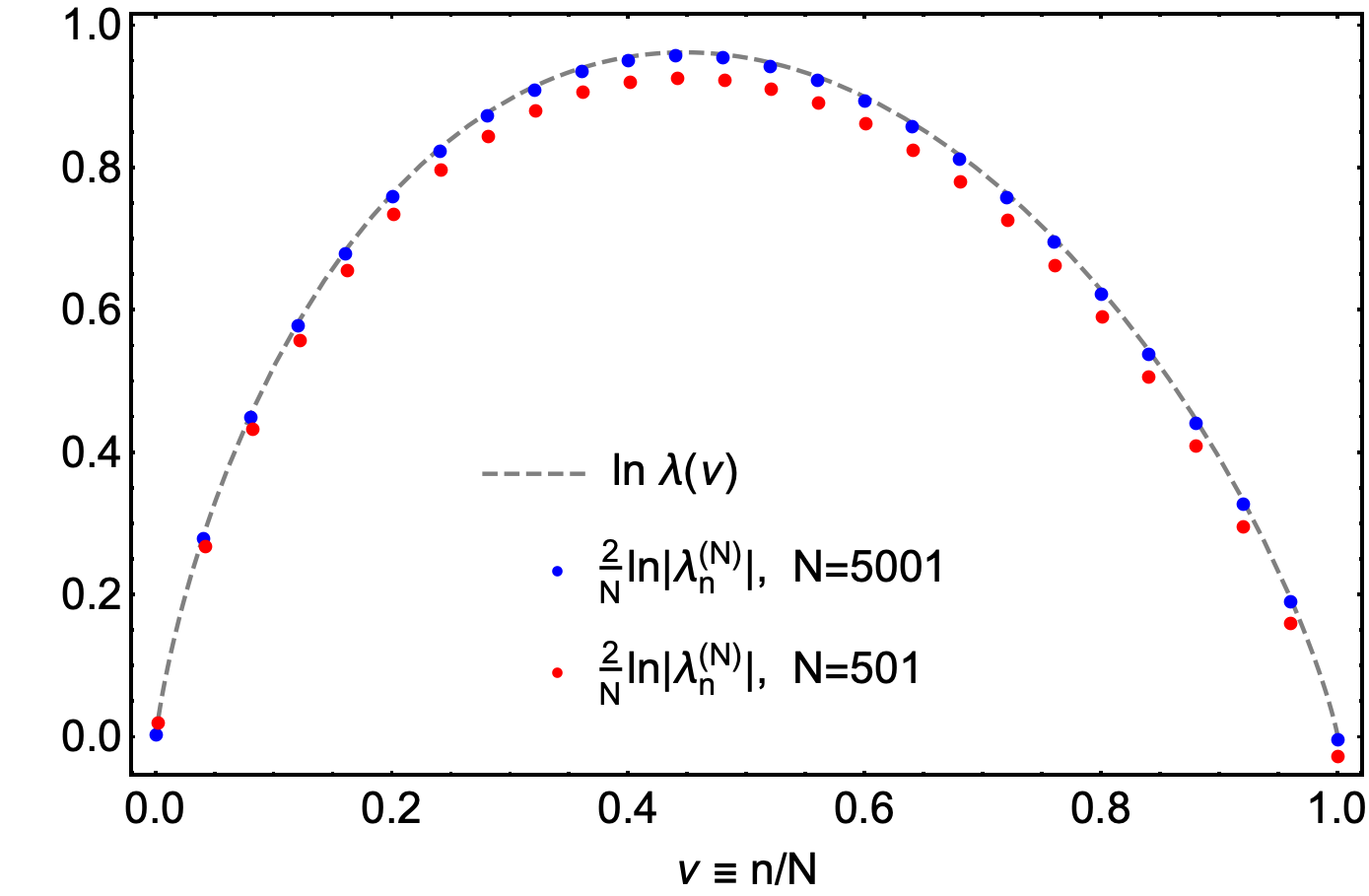}\label{fig:lambda_nu_comp}}
\caption{Comparison of the asymptotic function $\lambda(\nu)$ (dashed line) with the scaled explicit coefficients $\frac{2}{N}\ln|\lambda^{(N)}_n|$ (points) for initial wave amplitudes (a) $A_0=1$ and (b) $A_0=2$. This demonstrates the rapid convergence of the NKG potential coefficients to their large-$N$ limit.}
\end{figure}

\begin{figure}[h!]
\centering
{\includegraphics[width=0.5\textwidth]{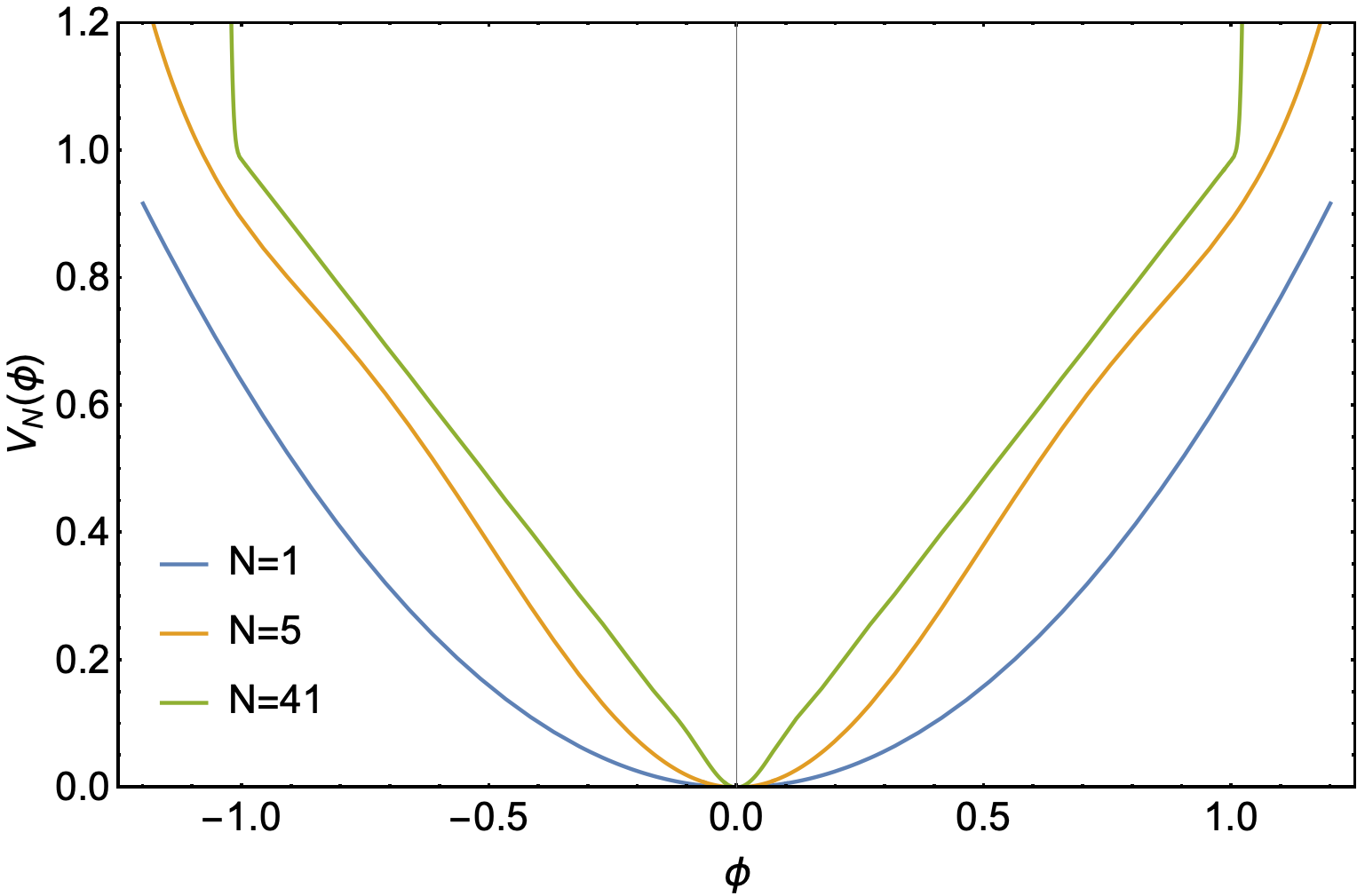}}
\caption{The potential $\eqref{eq:NKGpotential}$ is plotted with an initial amplitude $A_0=1$ for truncation parameters ${N=1}$, ${N=5}$, and ${N=41}$. The effect of including higher-order coupling constants (increasing $N$) on the potential shape is clearly visible.}
\label{fig:V_N}
\end{figure}

Figures $\ref{fig:lambda_nu2_comp}$ and $\ref{fig:lambda_nu_comp}$  provide a comparison between the asymptotic function $\lambda(\nu)$, defined by Equation $\eqref{eq:lambda_nu}$, and the scaled absolute values of the explicit coefficients $\frac{2}{N}|\lambda_n^{(N)}|$ derived in Equation $\eqref{eq:lambda_n}$. These figures demonstrate that as the truncation parameter $N$ increases (specifically for $N=\{501, 5001\}$), the explicit coefficients rapidly converge to the asymptotic function $\lambda(\nu)$. This convergence confirms that the truncated series offers an arbitrarily good approximation to the magnitude of the underlying field $|\phi|$ in the large-$N$ limit.

Figure \ref{fig:V_N} displays the potential expansion as a function of the number $N$. As $N$ increases, the potential's profile in the region $|\phi|<1$ increasingly resembles the V-shaped profile.

%%%%%%%%%%%%%%%%%%%%%%%%%%%%%%%%%%%%%%%%%%%

\subsubsection{Harmonics dispersion relation}
We have established the theoretical basis for the massless and quasi-massless SG propagation of Fourier modes. Given that the potential demonstrably generates harmonic wavelengths from an original input wavelength, $k_0$, under the condition $A_0 k_0^2\gg1$, a natural and requisite inquiry is: What are the corresponding frequencies of these propagating modes?

The initial configuration used in our previous analysis (Equation \ref{eq:ic_eigen}) does not explicitly define the relationship between the parameters ${k_0}$ (wavenumber) and ${\omega_0}$ (angular frequency). This relationship is inherently specified by the underlying physical model or, equivalently, by the governing field equation, which dictates the system's dispersion relation. For instance, in a linear KG model with mass $m$ (governed by a quadratic potential), an on-shell initial condition requires the standard dispersion relation:
$$\omega_0 = \sqrt{k_0^2 + m^2}.$$
Here, the sign of $k_0$ only determines the direction of wave propagation but does not impact the wave's frequency.
Since the perturbative mass ${m_0}$ for the input parameters $(A_0, k_0)$ has been calculated (see result \ref{eq:perturbativemass}), the dominant angular frequency ($\omega_0$) in the time domain for the SG system is consequently expected to follow a modified dispersion relation derived from the mass term:
\[
\omega_0 = \sqrt{k_0^2 + \frac{4}{\pi A_0}}.
\]

\begin{figure}[h!]
\centering
\subfigure[]{\includegraphics[width=0.49\textwidth, height=0.25\textwidth]{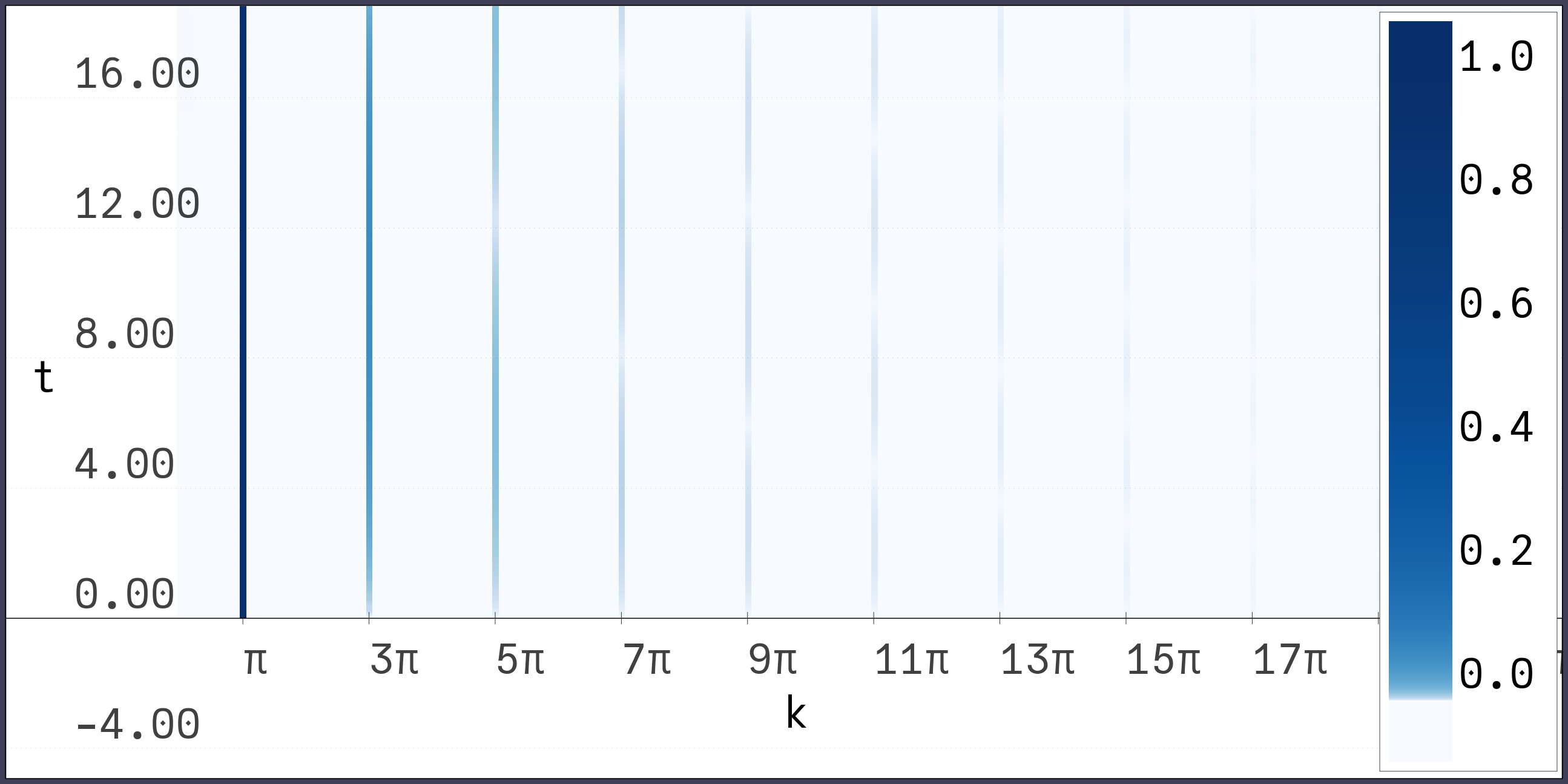}\label{fig:squaremodes_tx}}
\subfigure[]{\includegraphics[width=0.45\textwidth, height=0.25\textwidth]{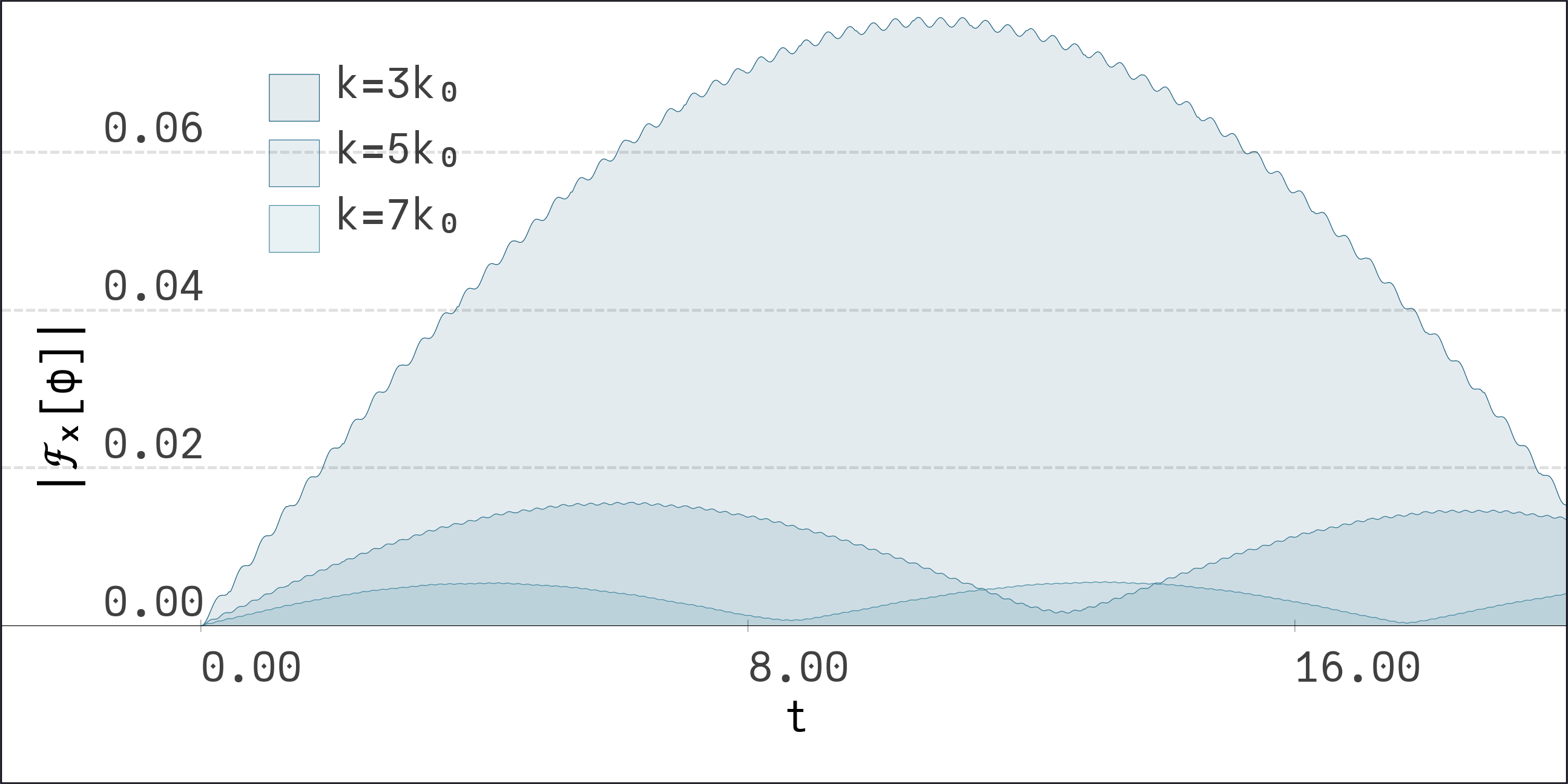}\label{fig:square_modes_profile}}
\caption{(a) Time evolution of the magnitude of the Fourier transform of the Sine-Gordon (SG) field's spatial component, $|\mathcal{F}_x[\phi](t,k)|$. (b) Temporal profiles of the magnitudes $|\mathcal{F}_x[\phi](k_n,t)|$ extracted from panel (a) at the harmonic wavenumbers $k_n=\{3k_0, 5k_0, 7k_0\}$, where $k_0=\pi$.}
\label{fig:squaremodes_profile}
\end{figure}

\begin{figure} 
	\centering
	\includegraphics[]{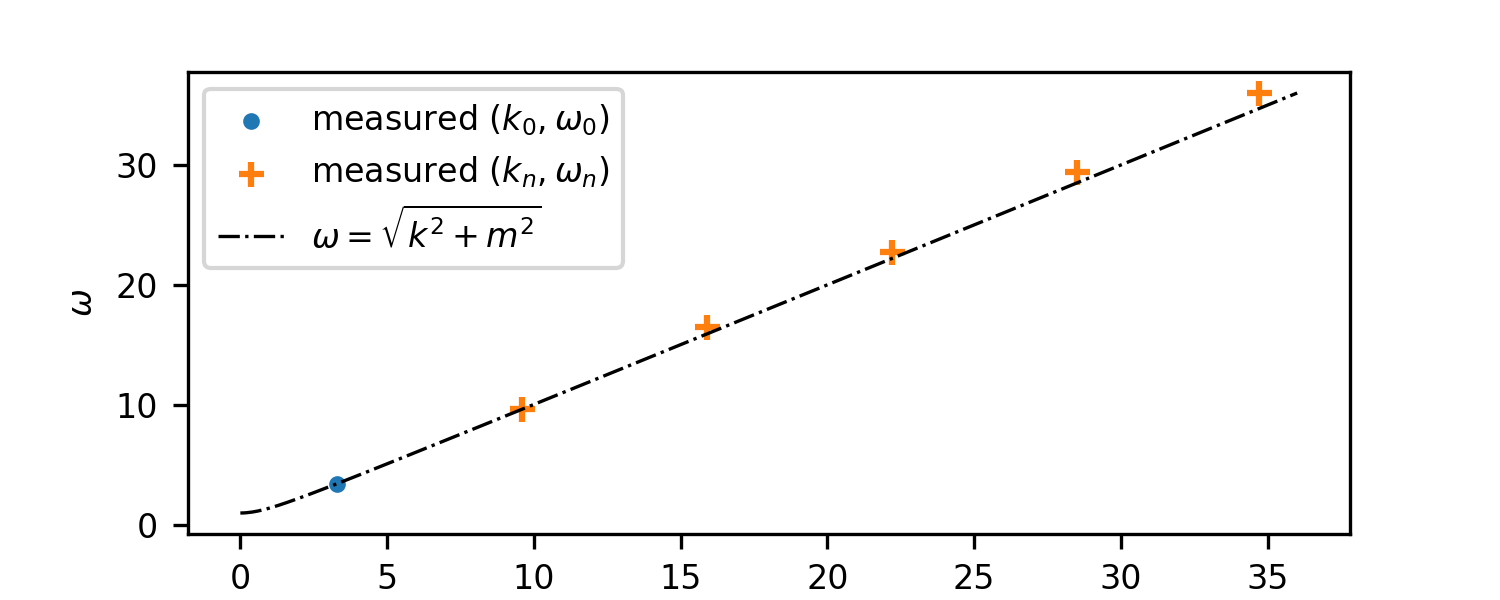}
	\caption[]{Dispersion relation illustrating the dominant mode ($n=1$, blue circle) and the five generated odd harmonics ($n=\{3,5,7,9,11\}$, light crosses). These data points were obtained via a two-dimensional Fourier transform in the time-space domain. The dot-dashed line represents a standard Klein-Gordon dispersion relation with a mass $m=1$.}
	\label{fig:generated_dispersion}
\end{figure}

Indeed, an initial configuration satisfying this modified dispersion relation leads to very stable wave propagation. The generated harmonics, $k_n$, appear to follow an approximately linear dispersion relation with respect to their corresponding angular frequencies, $\omega_n \simeq k_n$. The higher-order modes $k_n = n\pi$ (where $n=1, 3, 5, \dots$) are generated through the perturbative effects of the nonlinear term (Equation \ref{eq:fouriersigncos}). Figure \ref{fig:squaremodes_profile} illustrates the time evolution of these modes. By performing a time-domain discrete Fourier transform on the data presented in Figure \ref{fig:squaremodes_tx} and identifying the dominant amplitudes for each $(k_n, \omega_n)$ pair (where $n=1, 3, 5, \dots$), we obtain Figure \ref{fig:generated_dispersion}. This figure graphically depicts the dominant mode along with the five generated odd harmonics.

\subsubsection{Mass measurement methods} \label{sec:mass_inference}

This section details a method for extracting information from a field's effective potential, specifically the mass term and higher-order corrections, directly from the numerical field dynamics. This approach shifts focus from previous perturbative methods to an analysis of the complete field equation.

The core of the method involves generating and analyzing monochromatic signals within the field and constructing an amplitude map, $A(k,\omega)$, in energy-momentum space. This map is central to identifying the system's dispersion branches and, consequently, inferring information about the effective field-dependent potential. This process is conceptually analogous to an inverse propagation problem.

We concentrate on two key approaches for measuring the effective mass: one involves analyzing a system initiated with a monochromatic field configuration under periodic boundary conditions, and the other entails analyzing the field's response to an external monochromatic drive (signal production).

\begin{enumerate}
\item
{\bf Initial field configuration $(k_0\to\omega)$ method.}

We establish an initial "quasi-on-shell" field condition using a monochromatic wave, as defined in equation (\ref{eq:ic_eigen}), and employ periodic boundary conditions. This selection of boundary conditions is utilized to simulate an infinite spatial domain and eliminate edge effects. The initial condition corresponds to controlling the input wavenumber $k$ and subsequently measuring the resultant frequency distribution $\{\omega\}$. Within this method, the field is initialized with a specific wavenumber $k$ and permitted to propagate for a defined duration. The evolution of the angular frequency $\omega$ and the distribution of the generated signal across frequencies are monitored by computing time Fourier modes over the propagation period.

When executed as a single simulation, this procedure yields the time-domain Fourier transform of the field at a chosen spatial location $x$, which we set to $x=0$. To construct the complete dispersion map, we execute numerous such simulations for a set of equally-spaced values of $k$ and then aggregate these results into the map. To further guarantee smoothness at the boundaries under periodic boundary conditions, the input wavenumbers are constrained such that $k=L/(2\pi n)$, where $L$ denotes the spatial length of the simulation domain and $n$ is an integer. This methodology, applied to both the KG and SG equations, facilitates the numerical confirmation of the expected dispersion relations and the extraction of information pertaining to the effective potential, including the mass term.

The amplitude spectra presented in Figures \ref{fig:kg_dispersion_numeric} and \ref{fig:sg_dispersion} were generated using this method. For each wavenumber $k$ (horizontal axis), a simulation is performed, evolved temporally, and a spatial Fourier transform of the final time step provides the wavelength distributions (vertical axis). Figure \ref{fig:kg_dispersion_numeric} displays the reference computation, which is the resulting map and inferred mass as a function of the input wavenumber $k$, obtained from the procedure applied to the KG equation. Figure \ref{fig:sg_dispersion_a} presents a comparable map for the SG field, while Figure \ref{fig:sg_dispersion_b} illustrates its principal modes superimposed on the KG dispersion relation for $m=1$. Their close correspondence, particularly for $k>\pi/2$, is consistent with the hypothesis of massive behavior for monochromatic waves in the signum-Gordon field.

\begin{figure}[h!]
\centering
\subfigure[]{\includegraphics[width=0.48\textwidth]{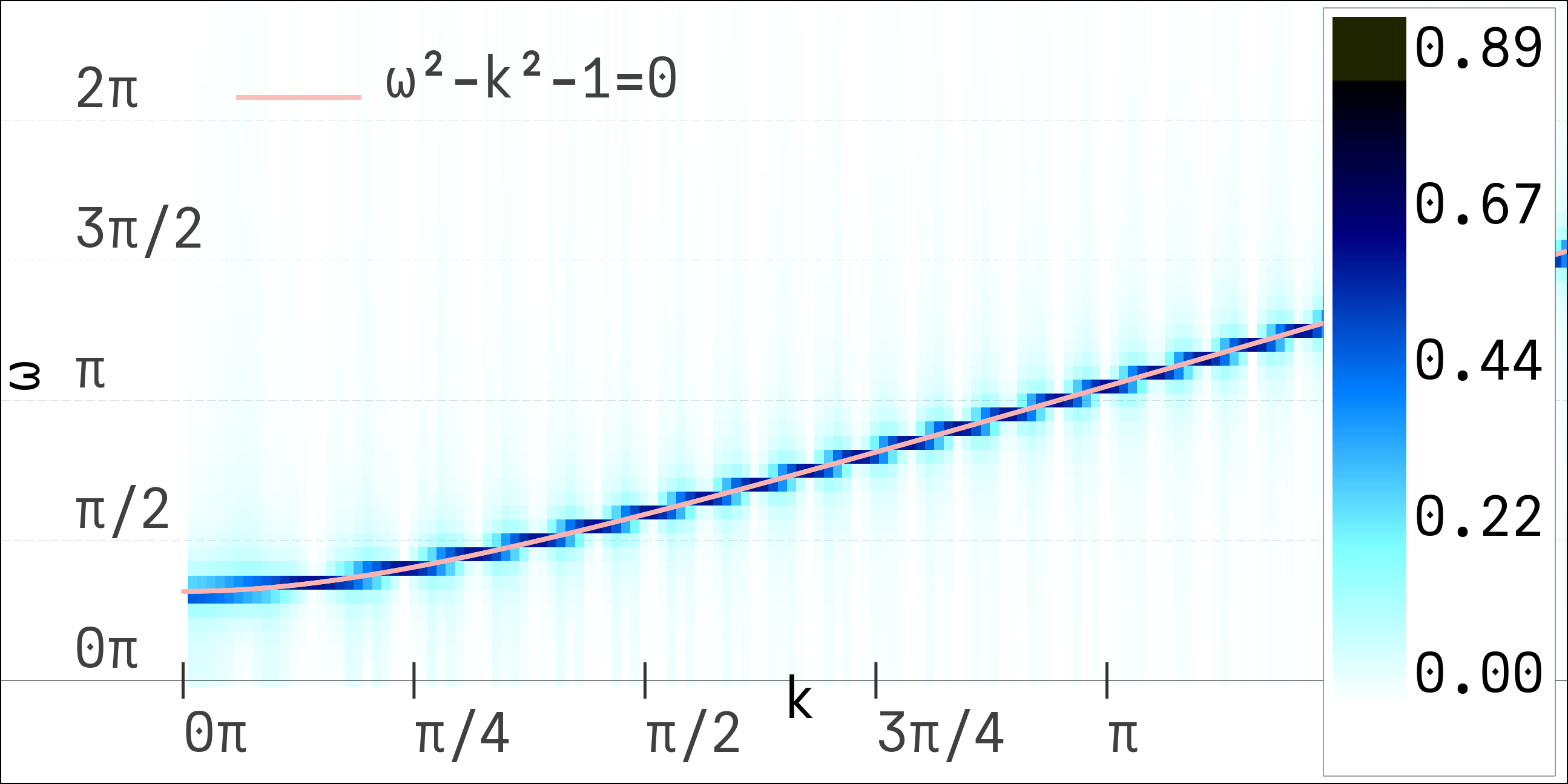}}\hskip0.5cm
\subfigure[]{\includegraphics[width=0.48\textwidth]{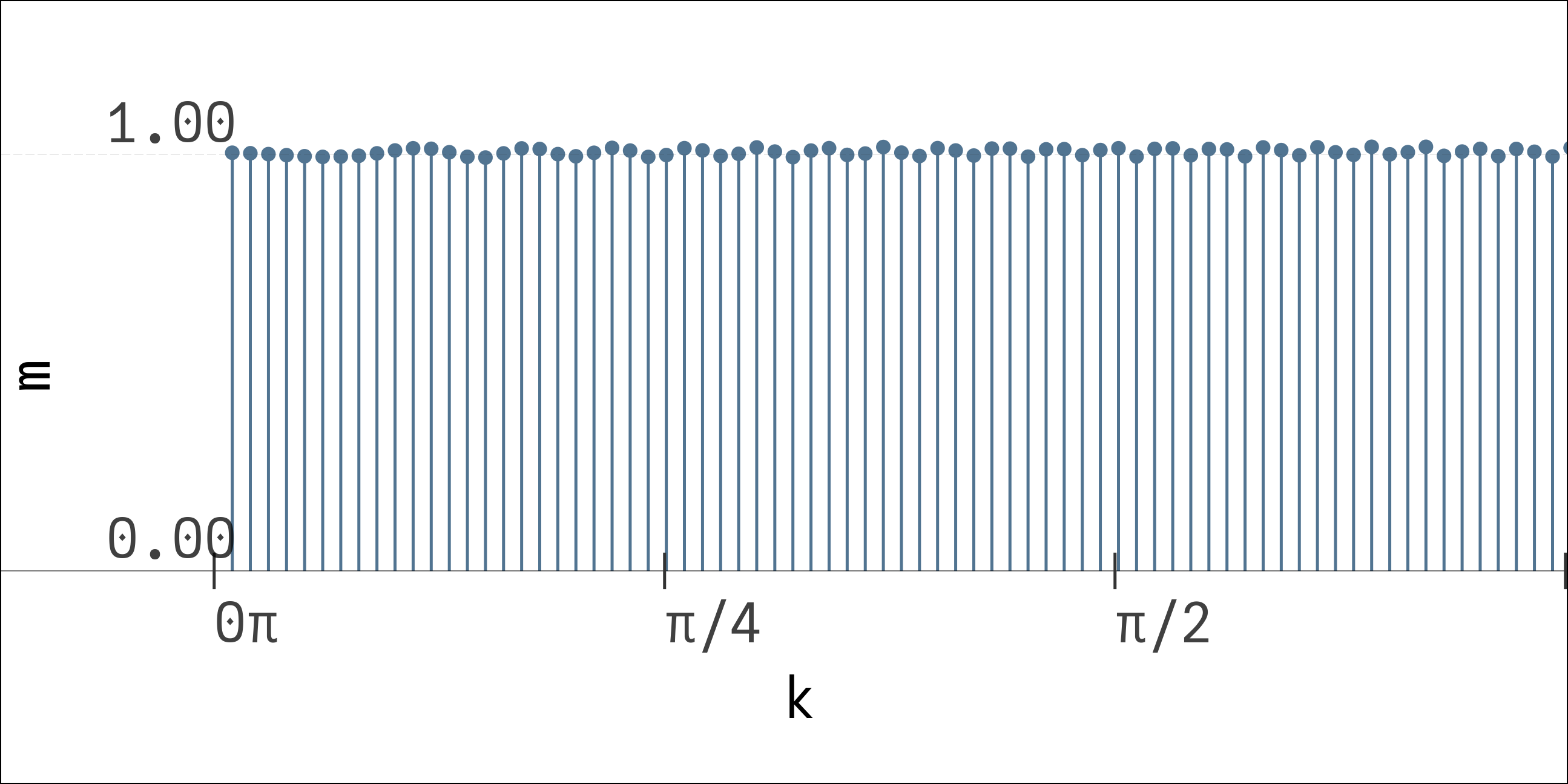}}
\caption{The ``on-shell'' KG disperision relation in energy momentum space is presented in Fig. (a), as computed by the signal production method, with the actual $m=1$ KG dispersion relation superimposed. Data was generated for the positive $k$ branch only, with KG mass set to $m=1$. Figure (b) shows the inferred field's mass $m$ from this data, which is, of course, constant (in this case $m=1$) for the linear KG field.}
\label{fig:kg_dispersion_numeric}
\end{figure}

\begin{figure}[h!]
\centering
\subfigure{\includegraphics[width=0.48\textwidth]{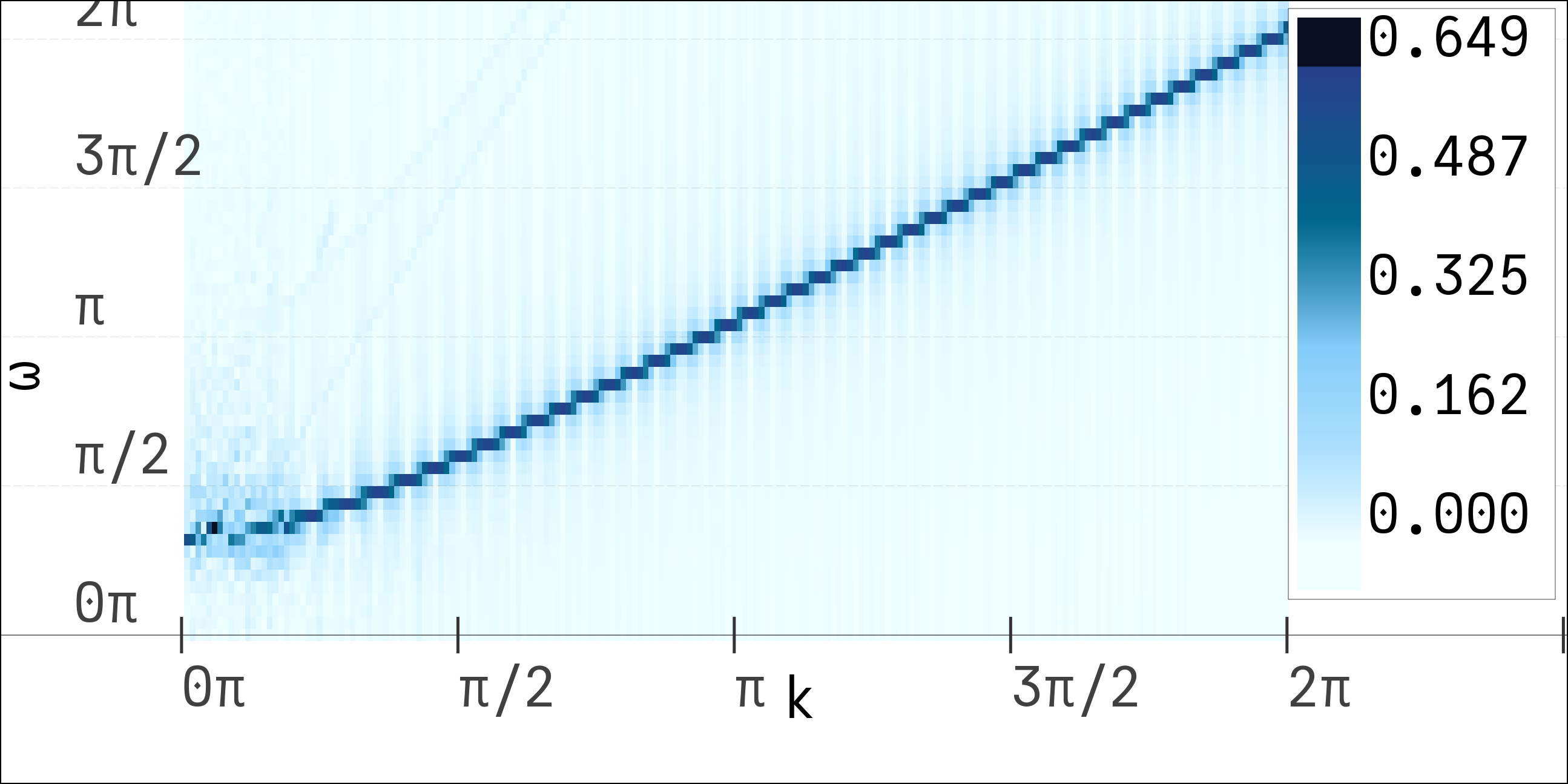}\label{fig:sg_dispersion_a}}\hskip0.5cm
\subfigure{\includegraphics[width=0.48\textwidth]{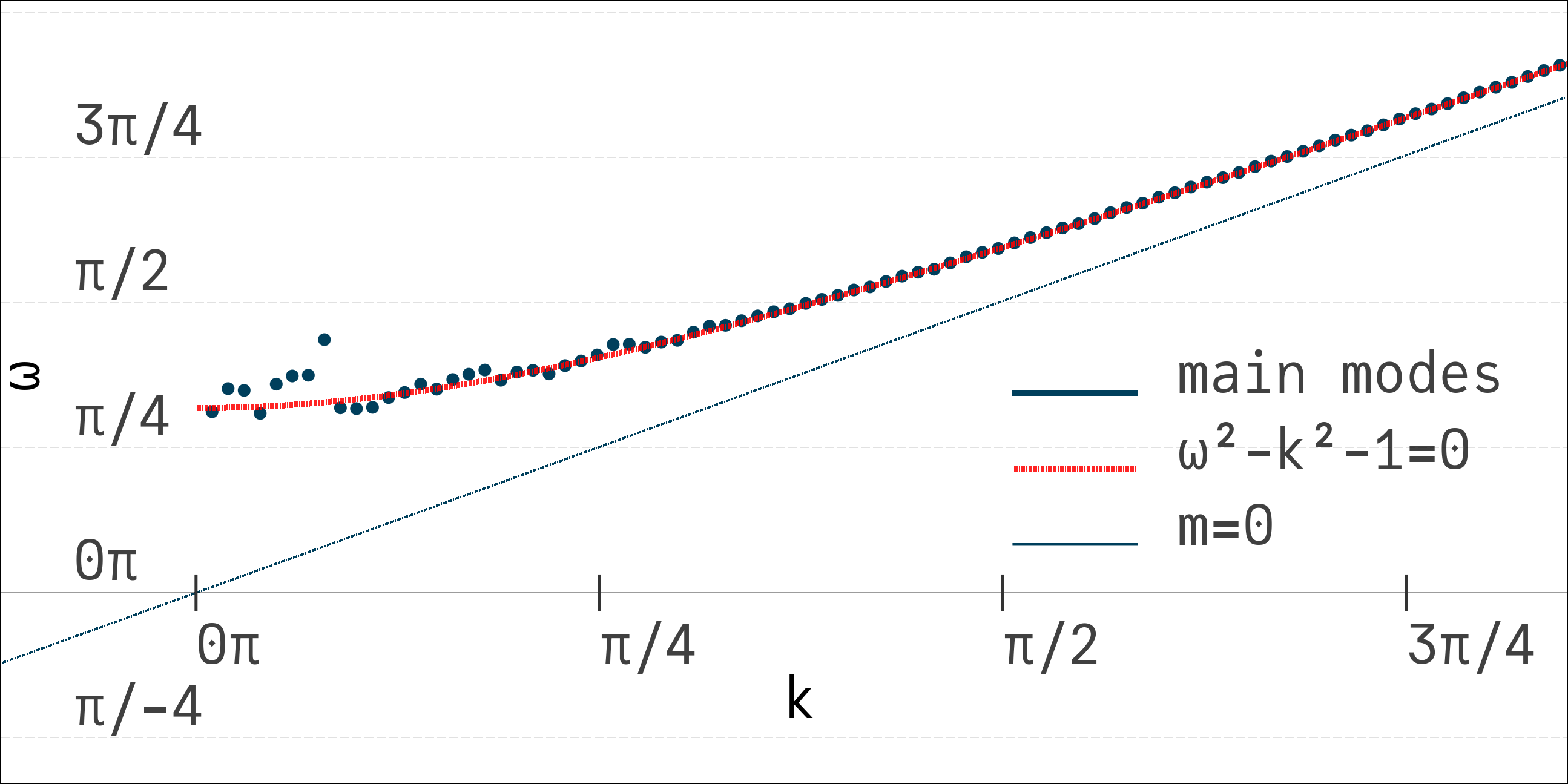}\label{fig:sg_dispersion_b}}\
\caption{SG dispersion relation for simulation with initial data $A=\dfrac{4}{\pi}\implies m=1$ and varying $k$. Figure (a) shows the a set of energy space of a simulated field in the timespan $t\in[0,40]$ (which for the simulation conditions is considered a short time). In Figure (b) the peak (main) modes have been extracted from the dispersion map in Figure (a) and plotted as bright dots, while the KG dispersion relation $\omega=+\sqrt{k^2+m^2}$ with $m=1$ is superposed to that and the same dispersion relation for $m=0$ is also shown for comparison. Notice the good quality of the matching, in special for $k>\pi/4$.}
\label{fig:sg_dispersion}
\end{figure}

\item
{\bf  Signal generated at $x=0$ ($\omega_0\to k$ method).}

This approach involves generating a monochromatic signal at one of the field's boundaries--for example, by imposing a specific time dependence on the field $\phi(t,0)$--and permitting the signal to propagate spatially up to a final time $t_f$.

Given an input frequency ${f_0 = \omega_0/2\pi}$, the propagating field is expected to select a suitable distribution of amplitudes as a function of wavenumbers, ${A(\omega_0, k) = A(k)}$. For instance, the theoretical result for the KG equation would be $A_{KG} = A_0\,\delta(\omega_0^2 - m^2)$, where $\delta(x)$ represents the Dirac delta distribution. This theoretical outcome indicates a sharp, predictable relationship between the input frequency and the resulting wavenumber.

The procedure is executed by applying a spatial Fourier transform, $\mathcal{F}_x$, to the field at the final time step to obtain the wavenumber distributions $A(\omega_0, k)$. Following the logic of the previous method, this process is repeated across a range of input frequencies $\omega_0$ to construct the complete amplitude map ${A(k, \omega_0)}$. Figure \ref{fig:sg_dispersion_signal_a} presents this resultant amplitude map. Figure \ref{fig:sg_dispersion_signal_b} displays the extracted magnitude peaks, $\text{max}[A(\omega_0, k)]$, plotted alongside two reference lines: the positive branch of the massive Klein-Gordon dispersion relation ($m=1$) and the massless dispersion relation ($m=0$).

\begin{figure}[h!]
\centering
{\subfigure{\includegraphics[width=0.48\textwidth]{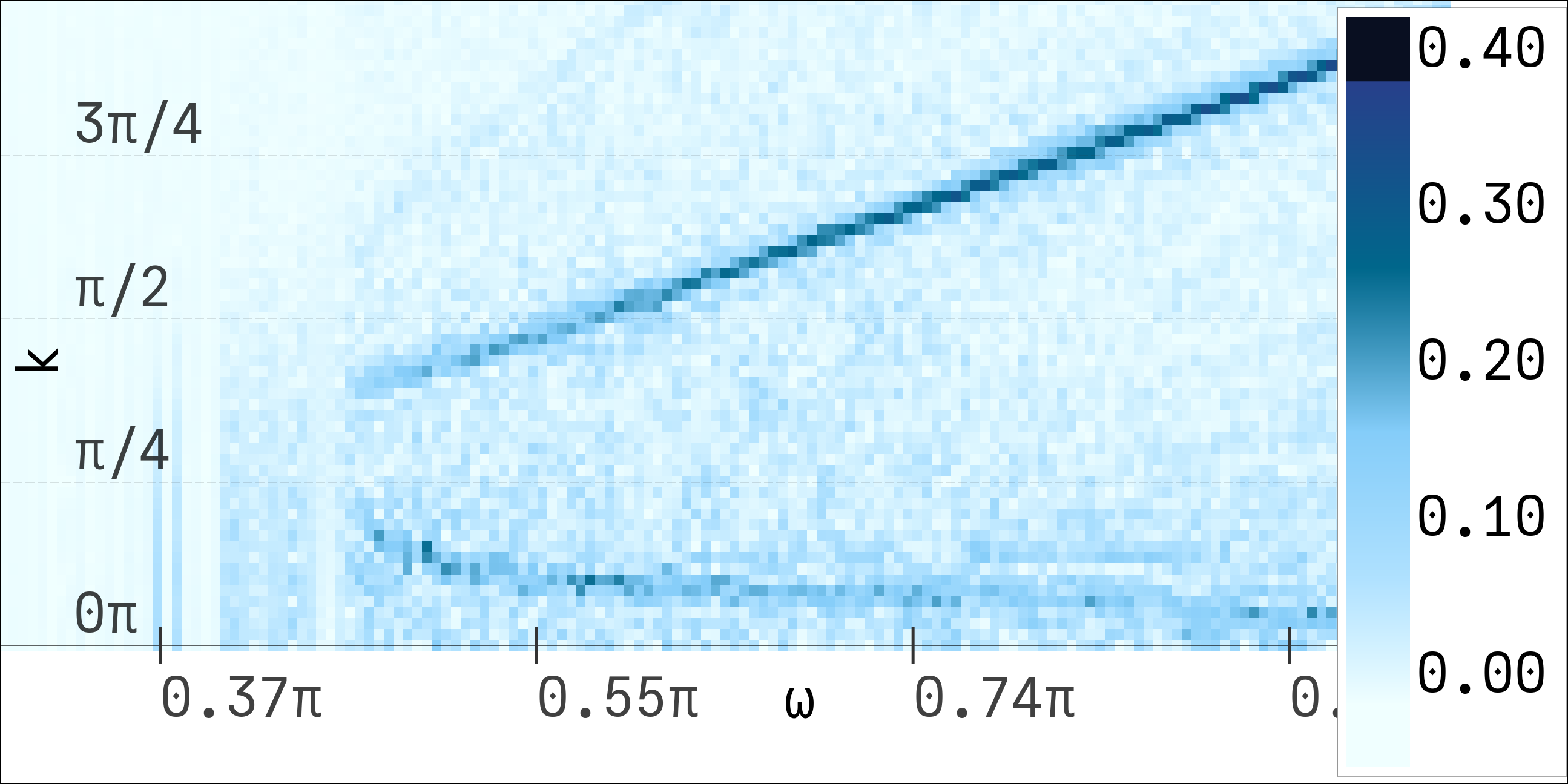}\label{fig:sg_dispersion_signal_a}}}\hskip0.5cm
{\subfigure{\includegraphics[width=0.48\textwidth]{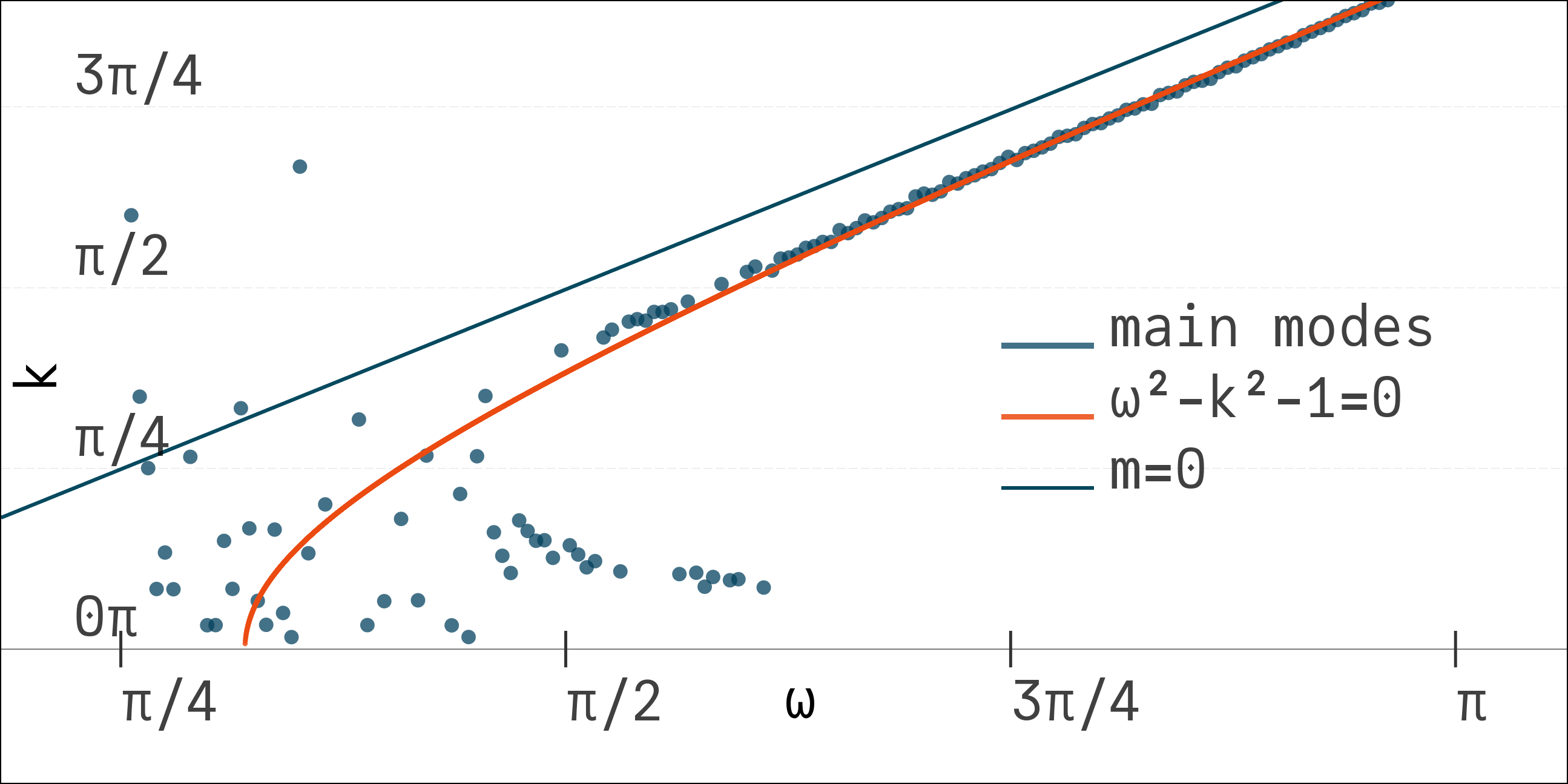}\label{fig:sg_dispersion_signal_b}}}\
\caption{The inferred amplitude spectrum of the Signum-Gordon (SG) field generated by boundary signals with an amplitude of $A_0=4/\pi$. This specific input amplitude results in an effective mass of $m_{\text{eff}}=1$. (a) The raw amplitude map ${A(k, \omega_0)}$ obtained using the $\omega_0 \to k$ method. (b) The extracted magnitude peaks (represented by circles) from the map in (a). These peaks are shown alongside the dispersion relations for the Klein-Gordon (KG) equation with a massive term ($m=1$) and a massless term ($m=0$) for reference.}
\label{fig:sg_dispersion_signal}
\end{figure}

\end{enumerate}

\section{Summary and remarks}

We have addressed the problem of mass in nonlinear scalar field models with non-analytic potentials, focusing on the SG model. This theory is nominally massless, as it lacks an explicitly given perturbative mass term in its potential. Because the potential's second derivative is ill-defined at the vacuum, the standard definition of mass is inapplicable. We introduced the concept of spectral mass by analyzing the field's dispersion relation through two complementary numerical methods: the initial field configuration ($k_0 \to \omega$) and the signal generation ($\omega_0 \to k$) approaches. By constructing amplitude maps in energy-momentum space, we tracked how the field selects specific frequencies and wavenumbers during its evolution.

Our analysis revealed that the propagation of wave trains depends critically on the product of the wave amplitude and the squared wavenumber. In the massless regime, where this product is large, the field follows an approximately linear dispersion relation. However, in the nonlinear regime, the SG potential acts as a source for nonlinear Fourier mode mixing, instantly generating higher odd harmonics. By establishing a formal mapping between the SG model and a truncated Nonlinear Klein-Gordon (NKG) model, we derived the leading analytical contribution to the spectral mass. This matching procedure allows us to identify the mass term directly from the potential's harmonic expansion. We demonstrated that a specific input amplitude results in an effective mass of unity, confirming that massive-like behavior can be induced in a nominally massless theory purely through nonlinear dynamics and the specific configuration of the initial wave.

The establishment of a spectral mass through nonlinear Fourier mode mixing opens several significant directions for future investigation. While this initial study was conducted in (1+1) dimensions, a critical extension involves studying scalar field models in higher spatial dimensions ($D>1$). In such cases, the scalar wavenumber $k$ is replaced by the vector $\vec{k}$, inherently increasing the complexity of the dispersion map $A(\vec{k}, \omega)$. This generalization is necessary to understand the role of spectral mass in driving the dynamics of higher-dimensional structures, such as the shock waves recently reported in the SG model in (2+1) and (3+1) dimensions \cite{Klimas:2023ife}.

Furthermore, the methodology should be extended to complex scalar field theories, such as the complex SG model or $CP^N$ models featuring V-shaped potentials. These models support compact structures like Q-balls and Q-shells; applying spectral analysis may elucidate the internal dynamics and interaction processes of these solutions. Key open questions remain regarding the long-term stability and persistence of the spectral mass -- specifically, whether the massive-like dispersion peak eventually damps toward a massless relation or remains a genuinely stable feature of the configuration.

Finally, representing the V-shaped potential through a finite harmonic decomposition provides a novel regularization of its non-analytic structure. By utilizing rapidly growing and alternating higher-order terms to compensate for the quadratic contribution, the resulting potential localizes infinite curvature at the origin while maintaining vanishing curvature elsewhere. This mimics the absolute value interaction and allows the model to be formulated in the functional integral as a Gaussian reference measure dressed by a local non-Gaussian weight, aligning with standard treatments of interacting scalar field theories.

\section*{Acknowledgements}
JSS would like to thank the support by CAPES (Coordena\c c\~ao de Aperfei\c coamento de Pessoal de N\'ivel Superior a Brasil (CAPES)) Scholarship during part of the production of this article. 
\bibliography{bibliography}

\end{document}